 \documentclass[twocolumn,trackchanges]{aastex631}
\usepackage{amsmath}
\usepackage{tabularx}
\usepackage{graphicx}
\usepackage{booktabs}       % professional-quality tables
\usepackage{amsfonts}       % blackboard math symbols
\usepackage{nicefrac}       % compact symbols for 1/2, etc.
\usepackage{microtype}
\usepackage{cleveref}
\usepackage{multirow}
\usepackage{makecell}
\usepackage{array}
\usepackage{subfigure}
\usepackage{stfloats}
\usepackage[utf8]{inputenc}

\renewcommand{\arraystretch}{1.2}
\shorttitle{Magnetic Neutron stars}
\shortauthors{I. A. Rather et al.}
\graphicspath{{./}{figures/}}
\begin{document}

\title{Heavy Magnetic Neutron Stars}

\author[0000-0001-5930-7179]{Ishfaq A. Rather}
\affiliation{Department of Physics \\
Aligarh Muslim University \\
Aligarh, 202002, India}

\author{Usuf Rahaman}
\affiliation{Department of Physics \\
	Aligarh Muslim University \\
	Aligarh, 202002, India}
%\collaboration{6}{(AAS Journals Data Editors)}

\author{V. Dexheimer}
\affiliation{Department of Physics, \\
	Kent State University, Kent OH 44242 USA}

\author{A. A. Usmani}
\affiliation{Department of Physics \\
	Aligarh Muslim University \\
	Aligarh, 202002, India}

\author{S. K. Patra}
\affiliation{Institute of Physics, Bhubaneswar 751005, India}
\affiliation{Homi Bhabha National Institute, \\
	Training School Complex, Anushakti Nagar,\\
	 Mumbai 400094, India}



\begin{abstract}
We systematically study the properties of pure nucleonic and hyperonic magnetic stars using a density-dependent relativistic mean field (DD-RMF) equations of state. We explore several parameter sets and hyperon coupling schemes  within the DD-RMF formalism. We focus on sets that are in better agreement with nuclear and other astrophysical data, while generating heavy neutron stars. Magnetic field effects are included in the matter equation of state and in general relativity solutions, which in addition fulfill Maxwell’s equations. We find that pure nucleonic matter, even without magnetic field effects, generates neutron stars that satisfy the potential GW190814 mass constraint; however, this is not the case for hyperonic matter, which instead only satisfies the more conservative 2.1 M$_{\sun}$ constraint.  In the presence of strong but still somehow realistic internal magnetic fields $\approx10^{17}$ G, the stellar charged particle population re-leptonizes and de-hyperonizes. As a consequence, magnetic fields stiffen hyperonic equations of state and generate more massive neutron stars, which can satisfy the possible GW190814 mass constraint but present a large deformation with respect to spherical symmetry.

\end{abstract}



\section{Introduction} \label{sec:intro}
Neutron stars (NSs) are considered to be the densest objects that are formed as a result of the gravitational collapse of supernovae. Their interior covers a wide range of densities (reaching $\sim$ 10 times the normal nuclear density) and, hence, invites the possibility of exotic degrees of freedom like quarks, hyperons, and kaons. They also provide a distinctive environment to investigate many unknowns in physics and astrophysics. The properties of the interior dense matter affect NS macroscopic properties, such as mass, radius, and the amount by which it can be deformed.

The most precise measurement of a NS mass is 1.4$M_{\sun}$, that of the Hulse-Taylor pulsar \citep{1975ApJ...195L..51H}. Recently, higher masses of binary neutron-star (BNS) systems like millisecond pulsar (MSP) PSR J1614-2230 (M= 1.97 $\pm$ 0.04 $M_{\sun}$) \citep{Demorest2010}, PSR J0348+0432 (M= 2.01 $\pm$ 0.04 $M_{\sun}$) \citep{Antoniadis1233232}, and PSR J0740+6620  (M= 2.04$^{+0.10}_{-0.09}$ $M_{\sun}$) \citep{Cromartie2020} have been measured with high precision. The gravitational wave detection by LIGO and Virgo Collaborations (LVC) of a BNS merger GW170817 event \citep{PhysRevLett.119.161101,PhysRevLett.121.161101} further provided a precise measurement of the NS mass, 1.16-1.60 $M_{\sun}$ for low spin priors and a maximum mass approaching 1.9 $M_{\sun}$ for high spin priors \citep{PhysRevX.9.011001}, which allowed one to study dense matter properties at extreme conditions due to the measurement of a new property of NSs, the tidal deformability. The latter provided a new constraint on the NS EoS and its variation with the radius approximated by the relation $\Lambda \propto R^5$ \citep{PhysRevD.81.123016,PhysRevC.95.015801} resulting in a strong constraint on the nuclear EoS at intermediate density. 

Very recently, a new gravitational wave event reported by LVC as GW190814 \citep{Abbott_2020a} was observed with a 22.2-24.3 $M_{\sun}$ black hole and 2.50-2.67 $M_{\sun}$ secondary component. The secondary component of GW190814 attracted a lot of attention as it contained no measurable tidal deformability signatures. As the mass of the secondary component of GW190814 lies in the lower region of the mass-gap (2.5$M_{\sun}$ $<$ M $<$ 5$M_{\sun}$), it raised the question whether it is a light black hole or a very massive NS. To understand this object, many interesting theories have been proposed recently regarding its nature as the most massive NS, lightest black hole, or fastest pulsar \citep{godzieba2020maximum,10.1093/mnrasl/slaa168,Zhang_2020,tsokaros2020gw190814,fattoyev2020gw190814,lim2020revisiting,tews2020nature,rather2021hadronquark,PhysRevD.102.041301,10.1093/mnras/staa1783,Biswas:2020xna,Roupas:2020nua,Zhou:2020xan,Wu:2021rxw,Khadkikar:2021yrj,Bombaci:2020vgw,Goncalves:2020joq,Christian:2020xwz,Demircik:2020jkc,Li:2020ias,Riahi:2020rkg,Dhiman:2007ck,Shahrbaf:2019vtf,Thapa:2020usm,Huang:2020cab,Lim:2020zvx,Tan:2020ics,Gupta:2019nwj,Dexheimer:2020rlp}.

% While the measurement of the NS mass is somewhat accurate (excluding GW190814 event), the precise measurement of the NS stellar radius does not exist yet.
Different values of NS radii have been obtained from the the analysis of X-ray spectra emitted by NSs \citep{PhysRevD.82.101301,Lattimer_2014,LATTIMER2016127,doi:10.1146/annurev-astro-081915-023322}. Lattimer \& Prakash  determined the range for radius at the NS canonical mass 10.7 $\le R_{1.4}$ $\le$ 13.1 km from the properties of symmetric matter and nuclear experiments \citep{LATTIMER2016127}. The recent measurement of the radius for the NS canonical mass inferred from PSR J0030+0451 by the Neutron Star Interior Composition
Explorer (NICER) \citep{2016SPIE.9905E..1HG} is R=13.02$_{-1.06}^{+1.24}$ km for M=1.44$_{-0.14}^{+0.15}$ $M_{\sun}$ \citep{Miller_2019a} or R=12.71$_{-1.19}^{+1.14}$ km for M=1.34$_{-0.16}^{+0.15}$ $M_{\sun}$ \citep{Riley_2019},  obtained using a Bayesian inference approach. In Ref. \citep{PhysRevLett.120.172702}, an upper limit on the NS radius for 1.4 $M_{\sun}$, $R_{1.4}$ $\le$ 13.76 km was deduced. The measurements from NICER provide an estimate of the canonical mass NS radius within a precision of 1 km. 
For a canonical mass NS with small radius, a softening of the pressure of  neutron matter at intermediate density is required, which results in a small value of the nuclear symmetry energy around saturation density $\rho_{0}$ \citep{2007PhR...442..109L,doi:10.1146/annurev-astro-081915-023322}. However, the requirement of the 2$M_{\sun}$ NS doesn't allow the pressure to be reduced too much. The small radius and large mass constraints are satisfied simultaneously by very few models that provide an additional accurate description of finite nuclei properties \citep{2015PhRvL.115p1101C,Jiang2015,Guichon:2018uew}.

At densities of a few times normal nuclear density, the composition of  matter inside NS is not known. With increasing density, the appearance of exotic degrees of freedom inside NSs is possible and has been studied over the past decade. In particular, the appearance of hyperonic matter under NS inner core conditions is energetically favored \citep{1960SvA.....4..187A,1985ApJ...293..470G}. The onset of hyperons reduces the pressure, leading to a softer EoS as they open new channels filling their Fermi sea, which decreases the maximum mass of NSs by about 0.5 $M_{\sun}$ \citep{GLENDENNING1982392,Chatterjee2016}. Several works has been performed recently considering hyperonic matter in NSs \citep{PhysRevC.100.015809,LI2018234,Li_2019,PhysRevD.102.041301,PhysRevC.95.065803,Li2018,Baldo:1999rq,Tolos:2016hhl,Colucci:2014wda,Gusakov:2014ota,Bhowmick:2014pma,Oertel:2014qza,Chatterjee:2015pua,Lopes:2017ovz,Isaka:2017nuc,Spinella:2018dab,Guichon:2018uew,Vidana:2018bdi,Fortin:2020qin,Tolos:2020aln,Gomes:2014aka,Dexheimer:2008ax,Fortin:2021umb}.

Heavy NSs are expected to contain exotic matter in their interior, even if they are rotating fast \citep{PhysRevC.97.035207,Dexheimer:2020rlp}. Very massive and/or fast rotating stars could be the result of accretion, or even a previous stellar merger, both of which have been shown to enhance stellar magnetic fields \citep{Pons2019,10.1093/mnras/stx2508,PhysRevD.99.054027,PhysRevD.102.096025,Zhong:2020etx,Most:2019kfe,Ciolfi:2019fie,Giacomazzo:2014qba,Xue:2019nlf,Raynaud:2020ist,Sur:2020imd}.
With exceptionally high density, a magnetic field reaching $\approx$ 10$^9$ to 10$^{18}$ G \citep{1995A&A...301..757B,Cardall_2001,10.1093/mnras/stu2706} is attainable in massive NS centers. Among various classes of compact stars available, Anomalous X-ray pulsars (AXPs) and the Soft gamma repeaters (SGRs), usually called magnetars \citep{2015SSRv..191..315M,doi:10.1146/annurev-astro-081915-023329}, are considered to have surface magnetic field within the order 10$^{14}$-10$^{15}$ G \citep{Harding_2006,Turolla_2015}.  Fast radio bursts (FRBs) have also shown evidence of magnetars \citep{Margalit_2020,10.1093/mnras/staa1783,Beloborodov_2020,Levin_2020,Zanazzi_2020}. 

In the interior of the magnetars, the magnetic field cannot be measured directly, and hence only estimated using theoretical models. The absolute largest value of magnetic field  a star can possess in the interior (usually taken as an upper bound for the magnetic field) can be estimated using the relativistic version of the virial theorem \cite{1993A&A...278..421B}, for which the negative total energy implies that the magnitude of the gravitational potential energy must be greater than the magnetic field energy, which results in magnetic field $\approx$ 10$^{18}$ G. As discussed in Refs. \cite{1995A&A...301..757B,Cardall_2001}, assuming a poloidal configuration, the maximum allowed central magnetic ﬁeld that still fulfills Einstein's and Maxwell's equations is around a few times 10$^{18}$ G, the exact value being dependent on the equation of state. Similar limits exist in purely toroidal configurations \cite{Pili:2017yxd}. As discussed in Refs.~\cite{Dexheimer:2016yqu,Pili:2017yxd}, the magnitude of magnetic field does not increase much beyond one order of magnitude within a star, regardless of the equation of state or magnetic field configuration. This implies that, regardless of its feasibility, the star would have to present $\geq10\times^{16}$ G on the surface in order to reach a magnetic field $\geq10\times^{17}$ G or higher in the center. Most of the magnetar observations inferred magnetic fields of $\sim10\times^{15}$ G or below, the only exception being data from the source 4U 0142+61 for slow phase modulations in hard x-ray pulsations (interpreted as free precession) that suggests magnetic fields of the order of $10\times^{16}$ G \cite{Makishima:2014dua}. Ref.~\cite{DallOsso:2018dos} provides $10\times^{16}$ G as a low bound for the same source. To be strongly structurally changed by magnetic fields, either more sources have not yet been discovered because their magnetic fields do not fit in the magnetar model (used to infer magnetic fields through period-period derivative diagrams), their magnetic fields decay quickly due to misalignment of rotation and magnetic axes \cite{Lander:2018und,Kuzur:2019mir}, or these stars are not stable, being formed for example in previous mergers \cite{Price:2006fi,Giacomazzo:2014qba,Most:2019kfe}. It should be mentioned that magnetic fields $\geq10\times^{15}$ G could be produced by three-dimensional simulations of supernova explosions with main requirement to produce a strong field dynamo being sufficient angular momentum in the progenitor star \cite{Raynaud:2020ist}.
Previous works have studied the effect of the magnetic field on the NS Equation of state (EoS) and on the stellar properties like mass and radius \citep{PhysRevLett.78.2898,Broderick_2000,Rabhi_2008,PhysRevC.89.045805,PhysRevLett.79.2176,Aguirre:2019ivr,Gomes:2017zkc,Dexheimer:2011pz,Pili:2017yxd,Felipe:2007vb,Paulucci:2013eia,Casali:2013jka,Rabhi:2008je}. The presence of strong magnetic fields deviates neutron-star structures from spherical symmetry very strongly and, hence, the spherically symmetric Tolman–Oppenheimer–Volkoff (TOV) equations can no longer be applied for studying their macroscopic structure \citep{10.1093/mnras/stu2706,Gomes:2019paw}.

In the present work, we employ recently proposed density-dependent Relativistic Mean-Field (DD-RMF) parameter sets to study the properties of NSs. By reproducing different hyperon-hyperon optical potentials, the values of the hyperon couplings are obtained, using several different coupling schemes, which are then used to model hyperons in our calculations. The presence of hyperons lowers NS maximum masses (to a value still larger than 2 $M_{\sun}$) by softening the EoS, although the radius at the NS canonical mass remains insensitive to them. We further analyze the effect of magnetic fields on nucleonic and hyperonic matter by employing a realistic chemical potential-dependent magnetic field \citep{DEXHEIMER2017487}. The macroscopic stellar matter properties for the magnetic EoSs are obtained using the publicly available Language Objet pour la RElativit\'e Num\'eriquE (LORENE) library \citep{LORENE,1995A&A...301..757B,1993A&A...278..421B}, which solves the coupled Einstein-Maxwell field equations in order to determine stable magnetic star configurations. In this case, NSs become more massive, especially the ones containing hyperons in their interior.

The main motivation behind our work is to demonstrate that a possible measurement of a neutron star mass $\sim2.5$ M$_\odot$ does not necessarily rule out exotic degrees of freedom in its interior, as several works in the literature claim. In this regard, despite the fact that the GW190814 data has been available for about a year \cite{Abbott:2020khf},  our conclusions differ significantly from all other works on the subject that have been published. While some argue that if the secondary object in GW190814 is a neutron star, it either has no exotic degrees of freedom \cite{Li:2020ias,Sedrakian:2020kbi} or, if it does, it rotates extremely fast \cite{Dexheimer:2020rlp,PhysRevC.103.055814} or uses an unusual approach to model deconfined quark matter \cite{PhysRevD.103.063042,PhysRevD.103.083015,PhysRevD.102.034031,PhysRevLett.126.162702,Demircik_2021,cao2020gw190814}, We investigate the possibility of a star with a strong magnetic field inside. We investigate how different particle populations and nuclear interactions affect microscopic and macroscopic stellar properties, and we illustrate how, if the secondary object in GW190814 is a massive neutron star, we can learn about the dense matter inside them. 

The paper is organized as follows: in section \ref{sec:headings}, the density-dependent RMF model is presented and the inclusion of the magnetic field for the beta-equilibrium EoS is discussed. The various DD-RMF parameter sets, nuclear matter properties, and hyperon couplings are discussed in section \ref{para}. The nucleonic and hyperonic EoSs are described in subsection \ref{hyp}, along with spherical solutions for star matter properties. Subsection \ref{magneticfield} deals with the neutron and hyperon EoS with effects of magnetic fields, together with the discussion of their stellar properties. Subsection \ref{sub3} presents additional results produced assuming different hyperon couplings. We summarize our results and outline an outlook for the future in sec. \ref{summary}. Finally, the equations of motion and the expressions for the energy and baryon density in the presence of magnetic fields are shown in Appendix \ref{A}.

\section{Theory and Formalism}
\label{sec:headings}

%\subsection*{{\textbf A. Nuclear Matter}}
The Lagrangian density is usually the starting point in RMF theory: the baryons interact through meson exchanges, including  the scalar-isoscalar sigma $\sigma$, the vector-isoscalar $\omega$, and the vector-isovector $\rho$.  The Lagrangian density contains contributions from the baryonic octet, leptons and the mesons, followed by the interactions between them. Within the DD-RMF model, the coupling constants are density-dependent \citep{PhysRevLett.68.3408}, as shown in the following.

The DD-RMF Lagrangian density  is written as \citep{Huang_2020}
\begin{align}\label{lag}
\mathcal{L}_m & =\sum_{B} \bar{\psi}_{B} \Biggl\{\gamma_{\mu}\Bigg(iD^{\mu}-g_{\omega}(\rho_B)\omega_{\mu}-\frac{1}{2}g_{\rho}(\rho_B)\rho_{\mu}\tau\Bigg) \nonumber \\
&-\Bigg(M-g_{\sigma}(\rho_B)\sigma\Bigg)\Biggr\} \psi_{B} 
+\frac{1}{2}\Bigg(\partial^{\mu}\sigma \partial_{\mu}\sigma-m_{\sigma}^2 \sigma^2\Bigg) \nonumber \\
&-\frac{1}{4}W^{\mu \nu}W_{\mu \nu}
+\frac{1}{2}m_{\omega}^2 \omega_{\mu} \omega^{\mu}
-\frac{1}{4}R^{\mu \nu} R_{\mu \nu} 
+\frac{1}{2}m_{\rho}^2 \rho_{\mu} \rho^{\mu} \nonumber \\
&+ \sum_l \bar{\psi}_l (i\gamma_{\mu} D^{\mu}-m_l)\psi_l,
\end{align}
where $B$ sums over the baryon octet ($n,p,\Lambda, \Sigma^+, \Sigma^0, \Sigma^-, \Xi^0, \Xi^- $)  and $l$ over $e^-$ and $\mu^-$. $\psi_B$ and  $\psi_l$ represent the baryonic and leptonic Dirac fields, respectively. The mesonic tensor fields and covariant derivative  are defined as 
\begin{align}
W^{\mu \nu}&=\partial^{\mu}W^{\nu}-\partial^{\nu}W^{\mu},\nonumber \\
R^{\mu \nu}&=\partial^{\mu}R^{\nu}-\partial^{\nu}R^{\mu}, \nonumber \\
D^{\mu} &= \partial^{\mu}+eQA^{\mu}. 
\end{align}

The Lagrangian density for the pure electromagnetic part is written as
\begin{equation}
\mathcal{L}_f=-\frac{1}{16\pi} F_{\mu \nu}F^{\mu \nu},
\end{equation}
where, $F_{\mu \nu}$ is the electromagnetic field tensor, $F_{\mu \nu}$=$\partial_{\mu}A_{\nu}-\partial_{\nu}A_{\mu}$. Hence, the total Lagrangian density in the presence of a magnetic field is
\begin{equation}
\mathcal{L}=\mathcal{L}_m +\mathcal{L}_f.
\end{equation}

The density dependent coupling constants in the DD-RMF model are parametrized by the relation
\begin{equation}
g_{iB}(\rho_B) = g_{iB}(\rho_0) f_i(x),
\end{equation}
for $i=\sigma,\omega$. The function $f_i(x)$, where $ x=\rho_B/\rho_{0}$ with $\rho_0$ as the nuclear matter saturation density, is defined as
\begin{equation}\label{eq5}
f_i(x) = a_i \frac{1+b_i (x+d_i)^2}{1+c_i(x+d_i)^2}.
\end{equation}
For the $\rho$ meson, the density-dependent coupling constant is given by an exponential relation as
\begin{equation}
g_{\rho}(\rho_B)=g_{\rho}(\rho_0)exp[-a_{\rho}(x-1)].
\end{equation}
This parametrization has eight parameters, which are reduced to three using the constraint conditions from refs \citep{TYPEL1999331,rather2021hadronquark} to reproduce symmetric and asymmetric nuclear matter properties.

For the present case that includes all baryons from the octet, the NS chemical equilibrium condition between different particles are
\begin{align}\label{eq13}
\mu_n &=\mu_{\Sigma^0} =  \mu_{\Xi^0}, \nonumber \\
\mu_p &=\mu_{\Sigma^+}=\mu_n -\mu_e, \nonumber \\
\mu_{\Sigma^-}& =\mu_{\Xi^-}=\mu_n +\mu_e,\nonumber \\
\mu_{\mu}&=\mu_e.
\end{align}
The charge neutrality condition follows as
\begin{equation}
\rho_p +\rho_{\Sigma^+}= \rho_e+\rho_{\mu^-}+\rho_{\Sigma^-}+\rho_{\Xi^-}.
\end{equation}

The expressions for the energy density and pressure in presence of a magnetic field can be obtained by solving the energy-momentum tensor relation
\begin{equation}
T^{\mu\nu} = T_m^{\mu\nu}+T_l^{\mu\nu}, 
\end{equation}
where \citep{PhysRevD.81.045015,PhysRevD.65.056001}
\begin{align}\label{eq11}
T_m^{\mu\nu} &= \mathcal{E}_m u^{\mu}u^{\nu}-P(g^{\mu\nu}-u^{\mu}u^{\nu}) \nonumber \\
&+\mathcal{M}B \Bigg(g^{\mu\nu}-u^{\mu}u^{\nu}+\frac{B^{\mu}B^{\nu}}{B^2}\Bigg), \nonumber \\
T_l^{\mu\nu} &= \frac{B^2}{4\pi}\Bigg(u^{\mu}u^{\nu}-\frac{1}{2}g^{\mu\nu}\Bigg)-\frac{B^{\mu}B^{\nu}}{4\pi}.
\end{align}
Here, $\mathcal{M}$ is the magnetization per unit volume and $B^{\mu}B_{\mu}=-B^2$.
For the nuclear matter in presence of a magnetic field, the single particle energy of all charged baryons and leptons is quantized in the direction perpendicular to the magnetic field. 

For a uniform magnetic field locally pointing in the $z$-direction, $B=B\hat{z}$, the total energy of a charged particle becomes
\begin{equation}
E_{cb}=\sqrt{k_z^2 +M^{*2}+2\nu|q|B},
\end{equation}
The quantity $\nu=\Big(n+\frac{1}{2}-\frac{1}{2}\frac{q}{|q|} \sigma_z\Big)=0,1,2,...$ indicates the Landau levels of fermions with electric charge $q$. $n$ is the orbital angular momentum quantum number and $\sigma_z$ is the Pauli matrix.
The Fermi momentum of all baryons charged $k_{F,\nu}^{cb}$ and leptons $k_{F,\nu}^l$ with Fermi energies $E_F^{cb}$  and $E_F^l$, respectively, are given as \citep{Broderick_2000}
\begin{align}
k_{F,\nu}^{cb}&=\sqrt{(E_F^{cb})^2-M_{cb}^{*2}-2\nu |q_{cb}|B},\nonumber \\
k_{F,\nu}^{l}& =\sqrt{(E_F^l)^2-m_l^{2}-2\nu |q_l|B}.
\end{align}
The highest value of $\nu$ is obtained with sum over Landau levels under the condition that the Fermi momentum of each particle is positive
\begin{align}
\nu_{max}&=\Bigg[\frac{(E_F^{cb})^2 -M_{cb}^{*2}}{2|q_{cb}|B}\Bigg], \nonumber \\
\nu_{max}&=\Bigg[\frac{(E_F^l)^2 -m_{l}^{2}}{2|q_{l}|B}\Bigg], 
\end{align}
for charged baryons $cb$ and leptons, respectively.

The expressions for the matter energy density and baryon density obtained in presence of magnetic field are shown in Appendix \ref{A}. From Eq. \ref{eq11}, the total energy density is
\begin{equation}
\mathcal{E}=\mathcal{E}_m +\frac{B^2}{8\pi}.
\end{equation}
The total pressure in the perpendicular and the parallel directions to the local magnetic field are
\begin{align}
P_{\perp}&=P_m-\mathcal{M}B +\frac{B^2}{8\pi},\nonumber \\
P_{\parallel}&=P_m-\frac{B^2}{8\pi},
\end{align}
where the magnetization is calculated as
\begin{equation}
\mathcal{M}=\partial P_m/\partial B.
\end{equation}

\section{Parameter Sets}
\label{para}
For the present study, we employ two recently proposed density dependent DD-MEX \citep{TANINAH2020135065}, and DD-LZ1 \citep{ddmex} parameter sets. These include the necessary tensor couplings of the vector mesons to nucleons and were obtained by fitting the ground state properties of finite nuclei, which considered parametric corrections and shell evaluations of mesons. Additionally, widely used  parameter sets DD-ME1 \citep{PhysRevC.66.024306} and DD-ME2 \citep{PhysRevC.71.024312} are also used.

Table \ref{tab1} displays the nucleon and meson masses and the coupling constants between nucleon and mesons for the given parameter sets. The  parameters $a,b,c,d$ for $\sigma$, $\omega$, and $\rho$ mesons are also shown.

\begin{deluxetable}{p{1.35cm}|p{1.4cm}p{1.4cm}p{1.4cm}p{1.42cm}}
	\tablenum{1}
	\tablecaption{Nucleon and meson masses and different coupling constants for various DD-RMF parameter sets. Nuclear matter properties: density, binding energy ($E/A$), incompressibility ($K_0$), symmetry energy ($J$), slope parameter ($L$), and nucleon effective masses at saturation density are also shown.\label{tab1}}
	\tablewidth{0pt}
	\tablehead{
		\colhead{} & \colhead{DD-LZ1} & \colhead{DD-ME1} & \colhead{DD-ME2} & \colhead{DD-MEX}} 

	\startdata
$m_n$ & 938.9000&939.0000&939.0000&939.0000\\
$m_p$&938.9000&939.0000&939.0000&939.0000\\
$m_{\sigma}$&538.6192&549.5255&550.1238&547.3327\\
$m_{\omega}$&783.0000&783.0000&783.0000&783.0000\\
$m_{\rho}$&769.0000&763.0000&763.0000&763.0000\\
$g_{\sigma}(\rho_0)$&12.0014&10.4434&10.5396&10.7067\\
$g_{\omega}(\rho_0)$&14.2925&12.8939&13.0189&13.3388\\
$g_{\rho}(\rho_0)$&15.1509&7.6106&7.3672&7.2380\\
\hline
$a_{\sigma}$&1.0627&1.3854&1.3881&1.3970\\
$b_{\sigma}$&1.7636&0.9781&1.0943&1.3350\\
$c_{\sigma}$&2.3089&1.5342&1.7057&2.0671\\
$d_{\sigma}$&0.3799&0.4661&0.4421&0.4016\\
$a_{\omega}$&1.0592&1.3879&1.3892&1.3926\\
$b_{\omega}$&0.4183&0.8525&0.9240&1.0191\\
$c_{\omega}$&0.5386&1.3566&1.4620&1.6060\\
$d_{\omega}$&0.7866&0.4957&0.4775&0.4556\\
$a_{\rho}$&0.7761&0.5008&0.5647&0.6202\\
\hline
\hline
$\rho_0(fm^{-3})$ &0.158&0.152&0.152&0.152\\
$E/A$(MeV)&-16.126&-16.668&-16.233&-16.140\\
$K_0$(MeV)&231.237&243.881&251.306&267.059\\
$J$(MeV)&32.016&33.060&32.310&32.269\\
$L$(MeV)&42.467&55.428&51.265&49.692\\
$M_n^*/M$&0.558&0.578&0.572&0.556\\
$M_p^*/M$&0.558&0.578&0.572&0.556\\
	\enddata
\end{deluxetable}

Some important nuclear matter properties for the DD-LZ1, DD-ME1, DD-ME2, and DD-MEX parameter sets are also shown in table (\ref{tab1}).
The symmetry energy parameter $J$ for the listed parameter sets is compatible with the values $J=31.6\pm 2.66$ MeV \citep{LI2013276}. The symmetry energy slope parameter $L$ also satisfies standard constraints $L=59.57\pm10.06$MeV \citep{PhysRevC.101.034303,DANIELEWICZ20141}. A very recent publication by the PREX collaboration reported a value for the neutron skin thickness of 208 Pb of $R_{skin}$ = (0.283 $\pm$ 0.071) fm \citep{PhysRevLett.126.172502}. Such a value is significantly larger than the $R_{skin}$ = 0.20 fm predicted by DD-ME1. Given the strong correlation between $R_{skin}$ and $L$, symmetry energy slope parameter values as large as $L$ $\approx$ 100 MeV have been proposed \citep{PhysRevLett.126.172503}. Although our parametrizations reproduce a lower symmetry energy and slope,  they are still in excellent agreement with a large amount of constraints, as shown in Fig.~1 from Ref.~\citep{Li:2021thg}. The currently accepted value of incompressibility determined from the isoscalar giant monopole resonance (ISGMR) lies in the range $K_0=240\pm20$MeV. The $K_0$ value for all the given parameter sets is within this range, except for the DD-MEX, which predicts a value little higher.

The density-dependent coupling constants of the hyperons to the vector mesons are determined from the SU(6) symmetry as \citep{PhysRevC.64.055805,PhysRevC.53.1416,Banik_2014, Tolos_2016}
\begin{align}\label{eq20}
\frac{1}{2}g_{\omega\Lambda}&=\frac{1}{2}g_{\omega\Sigma}=g_{\omega\Xi}=\frac{1}{3}g_{\omega N},\nonumber \\
\frac{1}{2}g_{\rho\Sigma}&=g_{\rho\Xi}=g_{\rho N}, g_{\rho \Lambda}=0. 
\end{align}
These couplings are calculated by fitting the hyperon optical potential obtained from the experimental data. See Refs.~\citep{Inoue:2019jme,Inoue:2018axd} for current lattice calculations and experimental data for these potentials.

The hyperon coupling to the $\sigma$ field is obtained, so as to reproduce the hyperon potential in symmetric nuclear matter (SNM) derived from the hypernuclear observables \citep{Banik_2014,RevModPhys.64.649} 
\begin{equation}
U_{\Lambda}^N(\rho)=g_{\omega \Lambda}\omega_0 +\sum_{R}-g_{\sigma \Lambda}\sigma_0. 
\end{equation}
For the present study, we reproduce the following optical potentials for the hyperons \citep{particles3040043}
\begin{align}\label{pot1}
U_{\Lambda}^N(\rho_0)&=-30 ~\mathrm{MeV},\nonumber \\
U_{\Sigma}^N(\rho_0)&=+30 ~\mathrm{MeV}, \nonumber \\
U_{\Xi}^N(\rho_0)&=-14 ~\mathrm{MeV}.
\end{align} 
These potentials correspond to the value of the density-dependent scalar couplings $g_{\sigma \Lambda}/g_{\sigma N} = 0.6105$, $g_{\sigma \Xi}/g_{\sigma N} = 0.3024$, and $g_{\sigma \Sigma}/g_{\sigma N} = 0.4426$.

\begin{figure}[htp!]
	\includegraphics[scale=0.32]{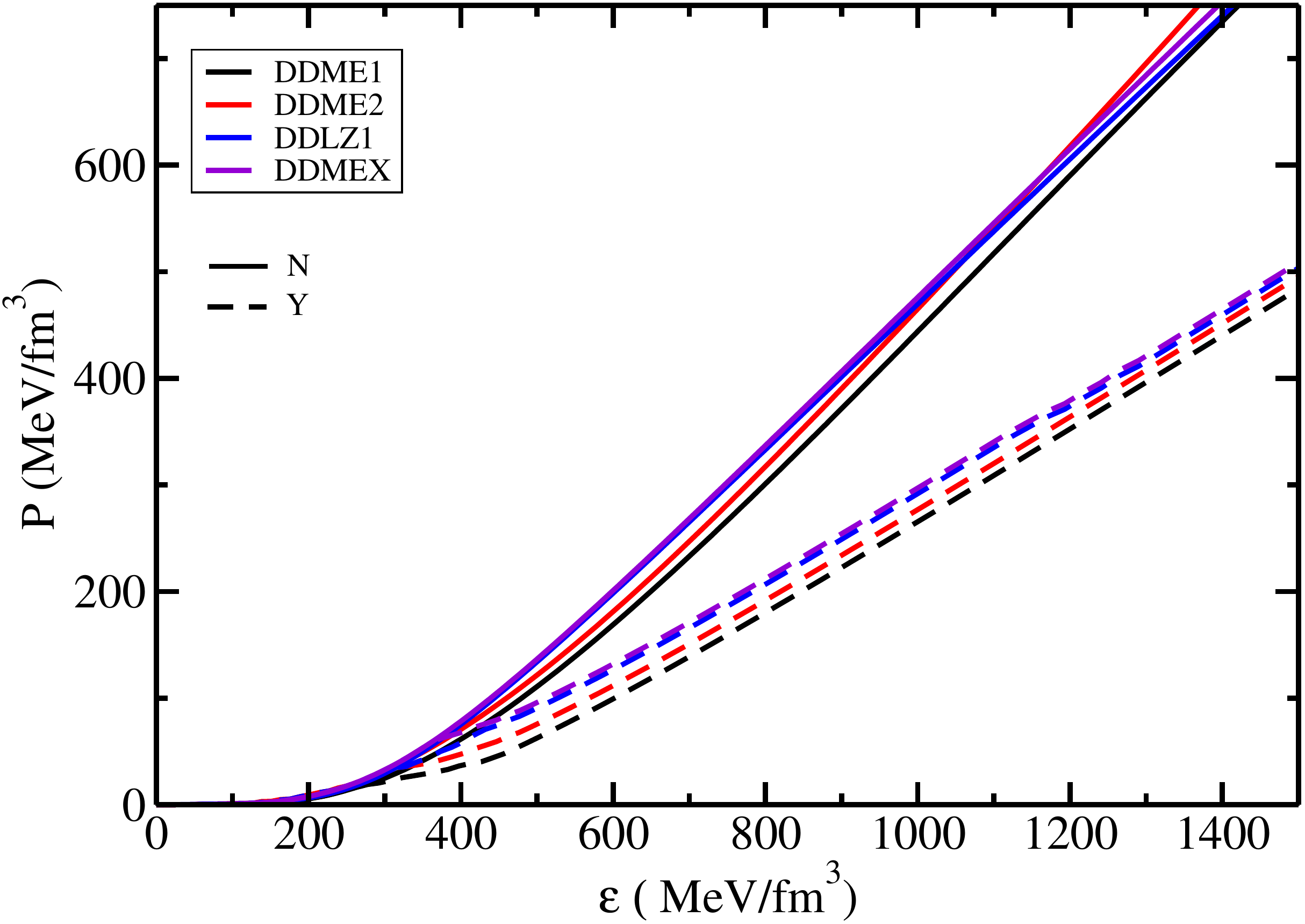}
	\caption{  Dense matter Equation of State for the DD-LZ1, DD-ME1, DD-ME2, and DD-MEX parameter sets. The solid lines represent pure nucleonic matter, while the dashed lines represent hyperonic matter including the entire baryon octet. }
	\label{fig1} 
\end{figure}

\section{Results and Discussions}
\label{results}

\subsection{Nucleonic and Hyperonic neutron stars}
\label{hyp}

The EoS of pure nucleonic (solid lines) and hyperonic matter (dashed lines) under chemical-equilibrium and charge neutrality conditions for several DD-RMF parameter sets are displayed in figure \ref{fig1}. For pure nucleonic matter, the DD-MEX parameter set produces a stiff EoS at low density region, while the DD-ME2 set produces a stiff EoS in the high density regime. The hyperons start to appear in the density range $\approx$ 300-400 MeV/fm$^3$ for all parameter sets. The onset of hyperonization softens the EoS (reduction in the pressure) due to the hyperons replacing the neutrons and opening new channels to distribute the Fermi energy. To build a unified EoS, the Baym-Pethick-Sutherland (BPS) EoS \citep{Baym:1971pw} has been used in the outer crust region. For the inner crust part, the EoS for non-uniform matter is generated by using DD-ME2 parameter set in a Thomas-Fermi approximation \citep{PhysRevC.79.035804,PhysRevC.94.015808,RATHER2021122189}. All the DD-RMF parameter sets considered are very similar in the outer crust density regime.

With the EoSs obtained for pure nucleonic and hyperonic matter, stellar matter properties like mass and radius are obtained by solving the Tolmann-Oppenheimer-Volkov (TOV) coupled differential equations
\citep{PhysRev.55.364,PhysRev.55.374} for a static isotropic spherically symmetric star
\begin{equation}\label{tov1}
\frac{dP(r)}{dr}= -\frac{[\mathcal{E}(r) +P(r)][M(r)+4\pi r^3 P(r)]}{r^2(1-2M(r)/r) },
\end{equation}
and
\begin{equation}\label{tov2}
\frac{dM(r)}{dr}= 4\pi r^2 \mathcal{E}(r),
\end{equation}
where $M(r)$ contained in the radius $r$ represents the gravitational mass for a specific EoS and a given choice of central energy density $\mathcal{E}_c$.
 
The tidal deformability $\lambda$ is defined as  \citep{PhysRevD.81.123016,PhysRevC.95.015801}
\begin{equation}\label{l1}
\lambda=-\frac{Q_{ij}}{\mathcal{E}_{ij}} = \frac{2}{3}k_2 R^5,
\end{equation}
where, $Q_{ij}$ represents the mass quadrupole moment of a star and $\mathcal{E}_{ij}$ corresponds to its external tidal field. The dimensionless tidal deformability $\Lambda$ is related to the compactness parameter $C=M/R$ as
\begin{equation}\label{l2}
\Lambda=\frac{\lambda}{M^5}=\frac{2k_2}{3C^5},
\end{equation}
where $k_2$ is the second love number. To estimate the love number along with the evaluation of the TOV equations, the function $y=y(R)$ as defined in ref \citep{Hinderer_2008} is  computed with the initial condition of $y(0)=2$ for the second love number. This value can be computed from the following first-order differential equations  \citep{PhysRevC.95.015801,Hinderer_2008} 
\begin{equation}\label{l4}
r\frac{dy(r)}{dr}+y(r)^2+y(r)F(r)+r^2 Q(r)=0,
\end{equation}
where,
\begin{align}\label{l5}
F(r)&=\frac{r-4\pi r^3 [\mathcal{E}(r)-P(r)]}{r-2M(r)},\\
Q(r)&=\frac{4\pi r\Big(5\mathcal{E}(r)+9P(r)+\frac{\mathcal{E}(r)+P(r)}{\partial P(r)/\partial\mathcal{E}(r)}-\frac{6}{4\pi r^2}\Big)}{r-2M(r)}\nonumber \\
&-4\Bigg[\frac{M(r)+4\pi r^3 P(r)}{r^2 (1-2M(r)/r)}\Bigg]^2.
\end{align}

Once the above equations are solved for $y=y(R)$ with $R$ being the radius of the spherical star in isolation, the second tidal love number is obtained from the following expression \citep{PhysRevD.81.123016}
\begin{align}\label{l3}
k_2&=\frac{8}{5}(1-2C)^2 [2C(y-1)]\Bigl\{2C(4(y+1)C^4
+(6y-4)C^3 \nonumber \\
&+(26-22y)C^2+3(5y-8)C-3y+6) \nonumber\\
&-3(1-2C)^2(2C(y-1)-y+2)\log\Big(\frac{1}{1-2C}\Big)\Bigr\}^{-1}.
\end{align}
Given the initial boundary conditions for the NS at the center $P(0)=P_c$, $M(0)=0$, and $y(0)=2$ with $P_c$ as the central pressure, to the surface of the star $P(R)=0$, $M(R)=M$, and $r(R)=R$, the above equations are solved for a given central density to determine the macroscopic properties of a NS.

\begin{figure}[htp!]
	\includegraphics[scale=0.32]{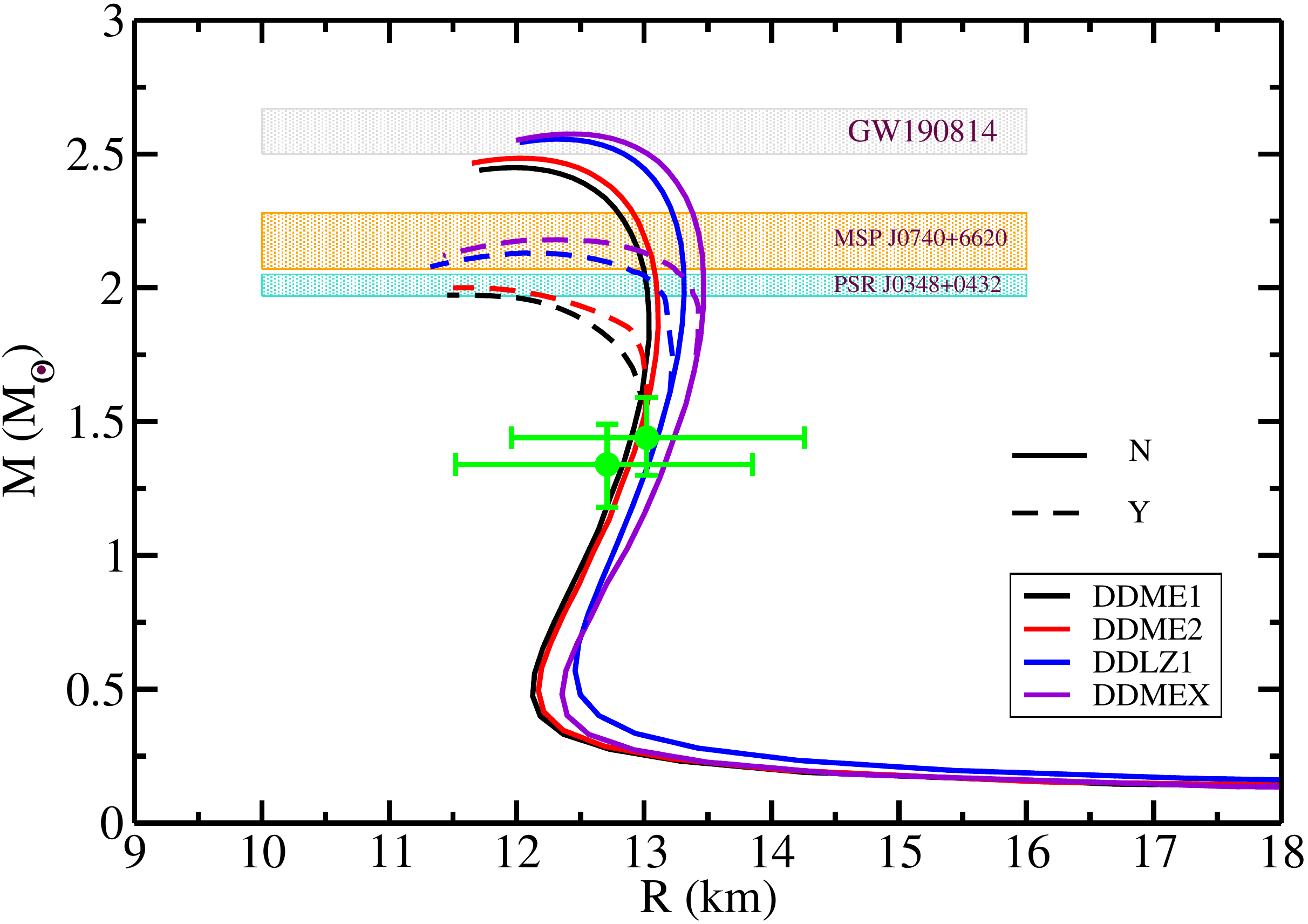}
	\caption{ Mass-Radius relation for pure nucleonic (solid lines) and hyperonic (dashed lines) stars using several DD-RMF parameters. The colored areas show recent constraints inferred from GW190814 and from the massive pulsars MSP J0740+6620 and PSR J0348+0432 \citep{Abbott_2020a,Cromartie2020,Antoniadis1233232}. The constraints on the mass-radius limits inferred from  NICER observations \citep{Miller_2019a,Riley_2019} are also shown.}
	\label{fig2} 
\end{figure}
Figure \ref{fig2} shows the mass-radius relation for pure nucleonic and hyperonic matter for the parameter sets DD-ME1, DD-ME2, DD-LZ1, and DD-MEX. The shaded areas represent the recent constraints on the NS maximum mass inferred from GW190814 (M=2.50-2.67 $M_{\sun}$), the massive pulsars MSP J0740+6620 (M=2.14$_{-0.09}^{+0.10}$ $M_{\sun}$), and PSR J0348+0432 (M=2.01$\pm 0.04$ $M_{\sun}$). The constraints on the radius limits around the NS canonical mass inferred from PSR J0030+0451 by NICER, R=13.02$_{-1.06}^{+1.24}$ km at M=1.44$_{-0.14}^{+0.15}$  $M_{\sun}$ \citep{Miller_2019a} and R=12.71$_{-1.19}^{+1.14}$ km at M=1.34$_{-0.16}^{+0.15}$ $M_{\sun}$ \citep{Riley_2019}, are also shown. For pure nucleonic matter, the DD-LZ1 and DD-MEX parameter sets reach a maximum mass of 2.55 and 2.57 $M_{\sun}$, with a radius of 12.30 and 12.46 km, respectively, indicating the possibility of GW190814 secondary component being a very massive NS. The hyperonic counterparts of the given DD-RMF parameter sets, which soften the EoS, produce a maximum mass of 2.18 and 2.13$M_{\sun}$ with a radius of 12.24 and 12.07 km respectively. For the DD-ME1 and DD-ME2 parameter sets, the NS maximum mass decreases from 2.45 and 2.48 to 1.98 and 2.01 $M_{\sun}$ respectively, while the respective radius decreases by $\approx$ 0.5 km. The radius at the NS canonical mass, R$_{1.4}$, remains insensitive to the appearance of hyperons. The hyperonic configurations satisfy the maximum mass limit from the massive pulsars MSP J0740+6620 and PSR J0348+0432, but are inconsistent with the GW190814  potential constraint (see discussion in  \citet{PhysRevD.102.041301}). All of the nucleonic and hyperonic configurations satisfy the mass-radius limits inferred from NICER.

\begin{figure}[htp!]
	\includegraphics[scale=0.32]{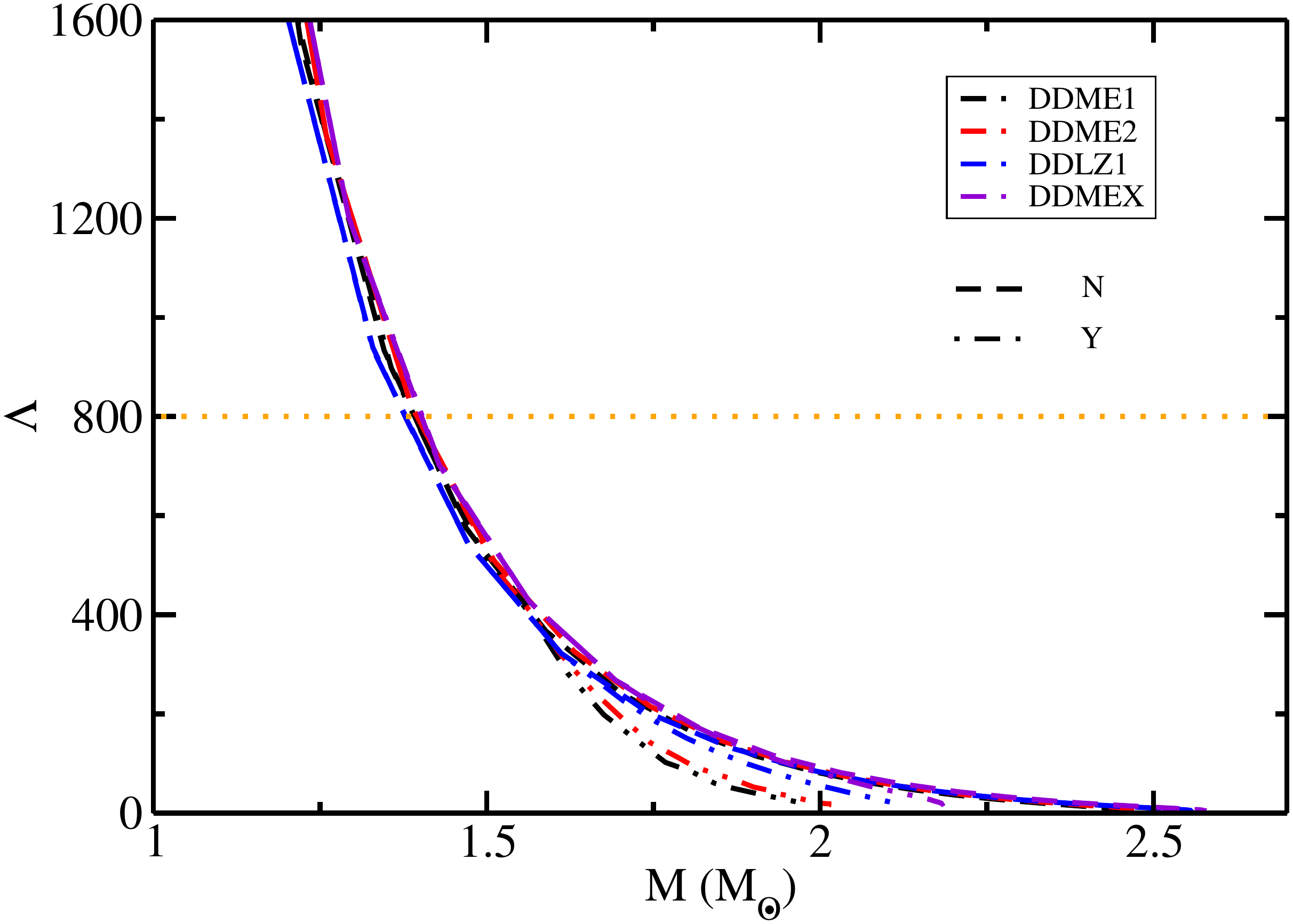}
	\caption{ Dimensionless tidal deformability variation with the NS mass for nucleonic (dashed lines) and hyperonic (dot dashed lines) stars using DD-LZ1, DD-ME1, DD-ME2, and DD-MEX parameter sets. The orange dotted line represents the upper limit on the dimensionless tidal deformability set by measurement from GW170817 \citep{PhysRevLett.119.161101}. }
	\label{fig3} 
\end{figure}

The variation of the dimensionless tidal deformability with the NS mass for the nucleonic and hyperonic stars based on the DD-RMF parametrizations are shown in figure \ref{fig3}.  The dimensionless tidal deformability of pure nucleonic as well as hyperonic matter for all the parameter sets lies well below the upper limit of $\Lambda_{1.4}$=800 obtained from the gravitational wave event GW170817 \citep{PhysRevLett.119.161101}. The shift in the tidal deformability for the hyperonic matter is seen as the mass increases. Table \ref{tab2} displays the different properties of neutron and hyperonic stars obtained with different DD-RMF parameter sets. A subsequent analysis by the LVC suggest a much smaller upper limit of 580 on the tidal deformability, which is smaller than all the values displayed in Table \ref{tab2} \cite{PhysRevLett.121.161101}. However, this value  corresponds to the 50\% confidence region, the 90\% confidence region extracted from the same data includes the values we reproduce.

\begin{deluxetable}{p{1.5cm}|p{1.7cm}p{1.5cm}p{1.5cm}
		p{1.5cm}|p{1.7cm}p{1.5cm}p{1.5cm}p{1.5cm}}
	\tablenum{2}
	\tablecaption{\hspace{8.3cm}{\bf{Table 2}}. Properties of pure nucleonic and hyperonic NSs for different DD-RMF parameter sets, including maximum mass,\newline 
	m\hspace{9.4cm} respective radius, and radius and dimensionless tidal deformability of a 1.4 $M_{\sun}$ star. \label{tab2}}
	\movetableright9.7cm	
	\tablewidth{0pt}
	\tablehead{
	&&\multicolumn{3}{p{2.5cm}|}{Nucleonic star} & %
	&\multicolumn{3}{p{2.5cm}}{Hyperonic star}\\
	\cline{2-9}
	&\parbox[t]{0.2cm}{\centering M$_{max}$ \\ ($M_{\sun}$) }  &\parbox[t]{0.2cm}{\centering R \\ (km) }&\parbox[t]{0.2cm}{\centering R$_{1.4}$  \\ (km) }&$\Lambda_{1.4}$&\parbox[t]{0.2cm}{\centering M$_{max}$ \\ ($M_{\sun}$) }  &\parbox[t]{0.2cm}{\centering R \\ (km) }&\parbox[t]{0.2cm}{\centering R$_{1.4}$  \\ (km) }&$\Lambda_{1.4}$ }
	\startdata 
	\hline
	DD-ME1 &2.449&11.981&12.898&689.342&1.983&11.515&12.898&689.342\\
	DD-ME2 &2.483&12.017&12.973&733.149&2.013&11.674&12.973&733.149\\
	DD-LZ1 &2.555&12.297&13.069&728.351&2.130&12.067&13.069&728.351\\
	DD-MEX &2.575&12.465&13.168&791.483&2.183&12.238&13.168&791.483\\
	\enddata
\end{deluxetable}

 \subsection{Magnetic stars}
\label{magneticfield}
%The ad hoc profile of the density-dependent magnetic field given by the relation \citep{PhysRevLett.79.2176}
%\begin{equation}
%B(\rho)=B_{surf}+B_{cent}\Bigl\{1-e^{\beta(\rho/\rho_0)^{\gamma}}\Bigr\},
%\end{equation}
%has been widely used to determine the structure of NS within the TOV system without solving the Maxwell's equations \citep{Tolos_2016,particles3040043,PhysRevC.89.015805,CHU2018447}. The parameters $\beta$ and $\gamma$ with typical values of (0.1,1) \citep{SINHA201343} and (0.0065,3.5) \citep{Tolos_2016} control the density at which magnetic field saturates and steepness the surface to core transition. The surface and core magnetic field with a value of around 10$^{15}$ G and 10$^{18}$ G respectively are strong enough to produce distinguishable effects on the NS properties. \\

In order to study the effects of magnetic fields on our microscopic description of matter, we employ a chemical potential-dependent magnetic field derived from the solutions of the Einstein-Maxwell's equations. The quadratic relation between the magnetic field and the chemical potential depends on the magnetic dipole moment and is given by \citep{DEXHEIMER2017487}
\begin{equation}
B^*(\mu_B)=\frac{(a+b\mu_B +c\mu_B^2)}{B_c^2} \mu,
\end{equation}
with $\mu_B$ being the baryon chemical potential in MeV and $\mu$ the dipole magnetic moment in units of Am$^2$ to produce $B^*$ in units of the electron critical field  $B_c=4.414\times 10^{13}$ G. The coefficients $a$, $b$, and $c$  taken as $a=-0.0769$  G$^2$/(Am$^2$), $b=1.20\times 10^{-3}$  G$^2$/(Am$^2$ MeV) and $c=-3.46\times 10^{-7}$  G$^2$/(Am$^2$ MeV$^2$) are obtained from a fit for the magnetic field in the polar direction of a star with a baryon mass of 2.2 $M_{\sun}$. 
%The magnetic field relation with density-dependent profile \citep{PhysRevLett.79.2176,PhysRevD.81.045015,Bandyopadhyay_1998} can lead to unphysical discontinuity in the presence of a phase transition like the phase transition to quark matter where the density changes discontinuously.

\begin{figure}[htp!]
	\includegraphics[scale=0.35]{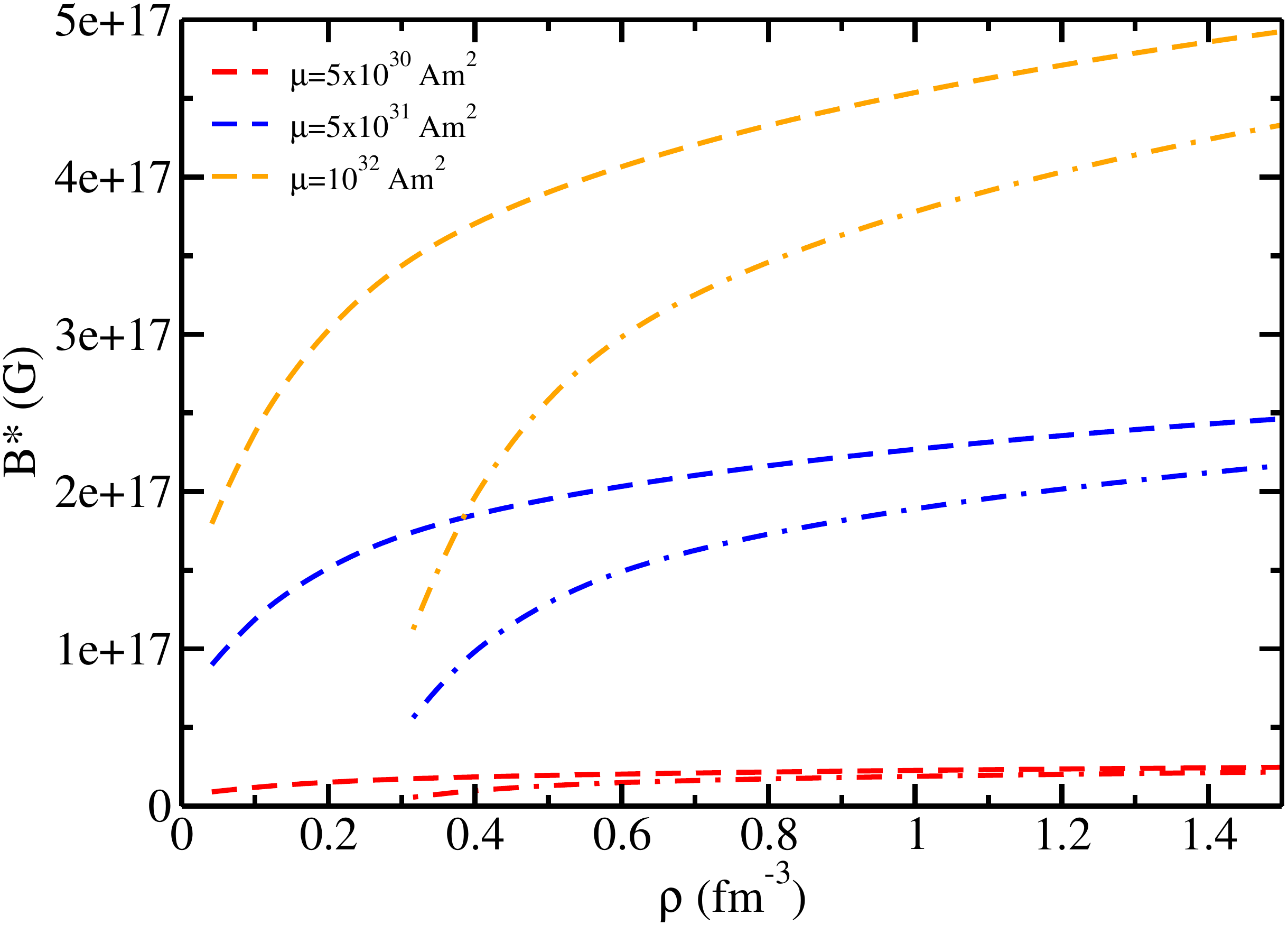}
	\caption{ Magnetic field profile as a function of baryon density for the DD-MEX EoS with different values of magnetic dipole moment. The dashed lines represent the profile for NS without hyperons, while the dot-dashed lines represent hyperonic stars.}
	\label{fig4} 
\end{figure}

Figure  \ref{fig4}  displays the magnetic field profile as a function of baryon density for a 2.2 $M_{\sun}$ baryonic mass star obtained for the DD-MEX EoS. The magnetic field effect on the DD-ME2 EoS has already been calculated \citep{particles3040043} and the DD-ME1 EoS predicts similar behavior. In Ref.~\citep{particles3040043}, spherically symmetric TOV equations were used for central magnetic fields $\approx$ 10$^{18}$ G with a density-dependent and universal profile for magnetic field. Since the DD-MEX EoS predicts a heavier NS than other parameter sets, we choose this parameter set to study magnetic effects and verify whether this model predicts the possibility of the GW190814 secondary component being a hyperonic magnetar. In fig. \ref{fig4}, the dashed curves represent a NS without hyperons and dot-dashed curves represent a NS with hyperons. It is clear that the magnetic field produced by the NS without hyperons is larger. This illustrates that the transition from $\mu_B$ to $\rho_B$ is model and particle population dependent. For magnetic dipole moment greater than 10$^{31}$ Am$^2$, the magnetic field produced at large densities is larger than 10$^{17}$ G, which is strong enough to cause a large deformation in the NS structure. The values of the magnetic field produced at the surface and at large densities using different values of the magnetic dipole moment for NSs with and without hyperons are shown in table \ref{tab3}. For a magnetic dipole moment 10$^{32}$ Am$^2$, the magnetic field produced at large densities is greater than 4$\times$ 10$^{17}$ for both cases.
\begin{figure}[htp]
	\includegraphics[scale=0.32]{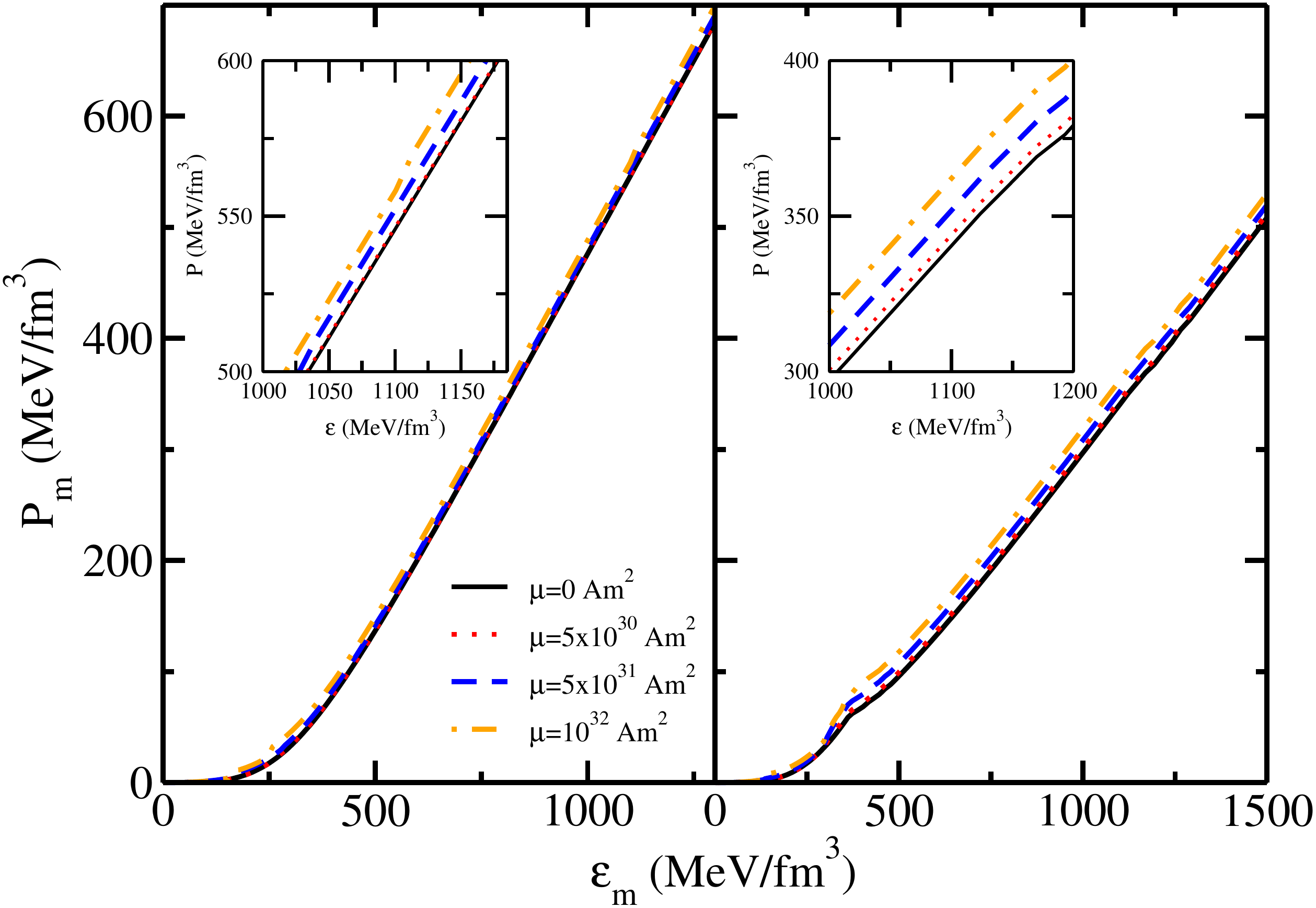}
	\caption{ Variation of matter pressure in transverse direction vs energy density for the DD-MEX parameter set without and with magnetic field effects at different values of magnetic dipole moment. Left panel depicts the EoSs without hyperons and right panel depicts the EoS with hyperons. The insets in each panel show the variation in the pressure at higher value of the energy density for different magnetic moments. }
	\label{fig5} 
\end{figure}
\begin{deluxetable}{p{1.35cm}|p{1.4cm}p{1.4cm}|p{1.4cm}p{1.42cm}}
	\tablenum{3}
	\tablecaption{Magnetic field at low densities B$_s$ (corresponding to stellar surfaces) and at high values of densities B$_c$ calculated for the DD-MEX EoS with and without hyperons.\label{tab3}}
	\tablewidth{0pt}
	\tablehead{
		&\multicolumn{2}{p{2.7cm}|}{Nucleonic star} & %
		\multicolumn{2}{p{2.1cm}}{Hyperonic star}\\
		\hline
	$\mu$ (Am$^2$)& B$_{s}$ (G)&B$_{c}$ (G)&B$_{s}$ (G)&B$_{c}$ (G)}
		\startdata
	5$\times$10$^{30}$ &1.01$\times$10$^{15}$&2.59$\times$10$^{16}$&6.65$\times$10$^{15}$&1.96$\times$10$^{16}$\\
	5$\times$10$^{31}$&8.98$\times$10$^{16}$&2.28$\times$10$^{17}$&5.83$\times$10$^{16}$&1.89$\times$10$^{17}$\\
	10$^{32}$&1.79$\times$10$^{17}$&4.55$\times$10$^{17}$&1.12$\times$10$^{17}$&3.77$\times$10$^{17}$ 
	\enddata
\end{deluxetable}

\begin{figure*}
	\gridline{\fig{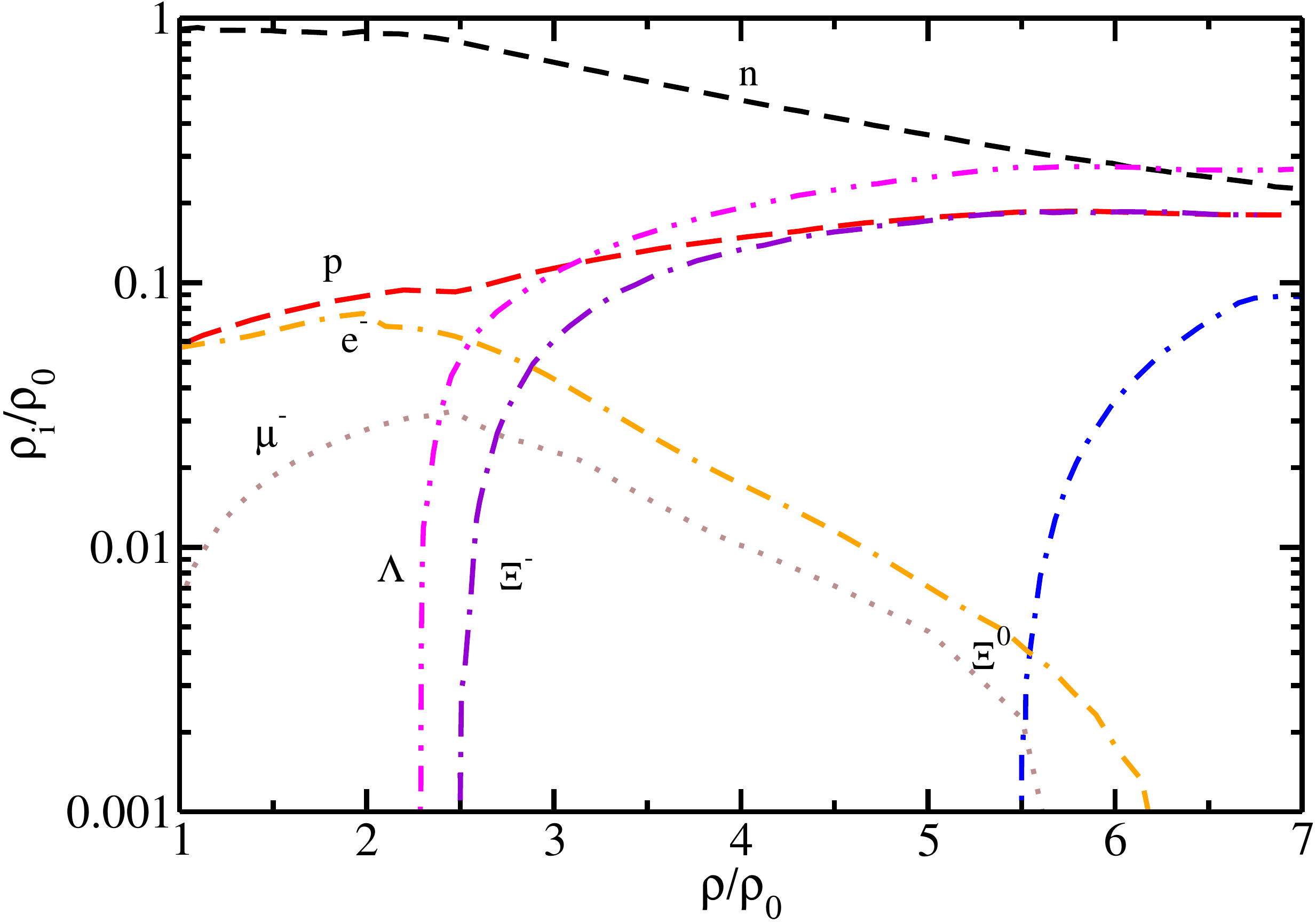}{0.5\textwidth}{(a)}
		\fig{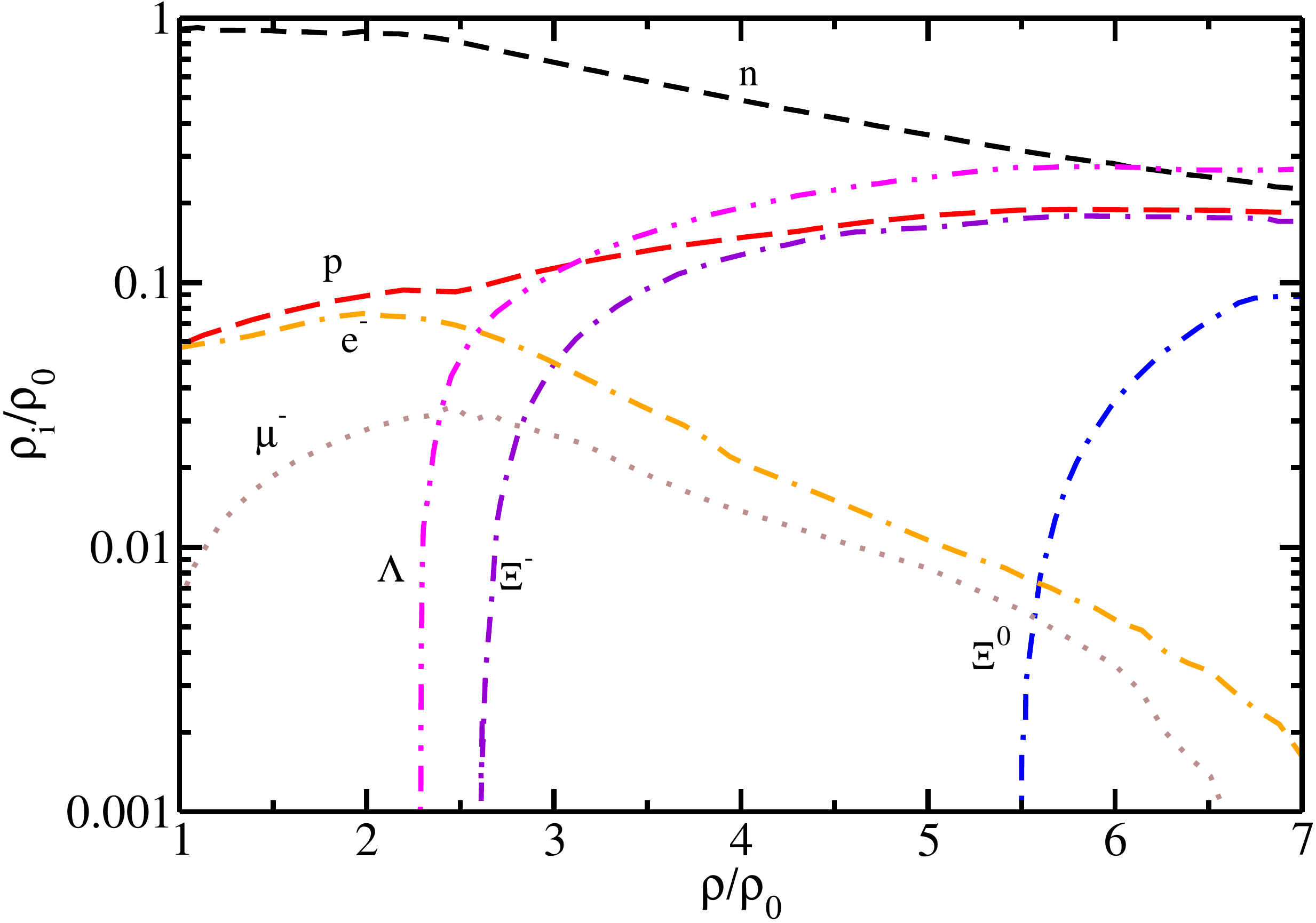}{0.5\textwidth}{(b)}
	}
	\gridline{\fig{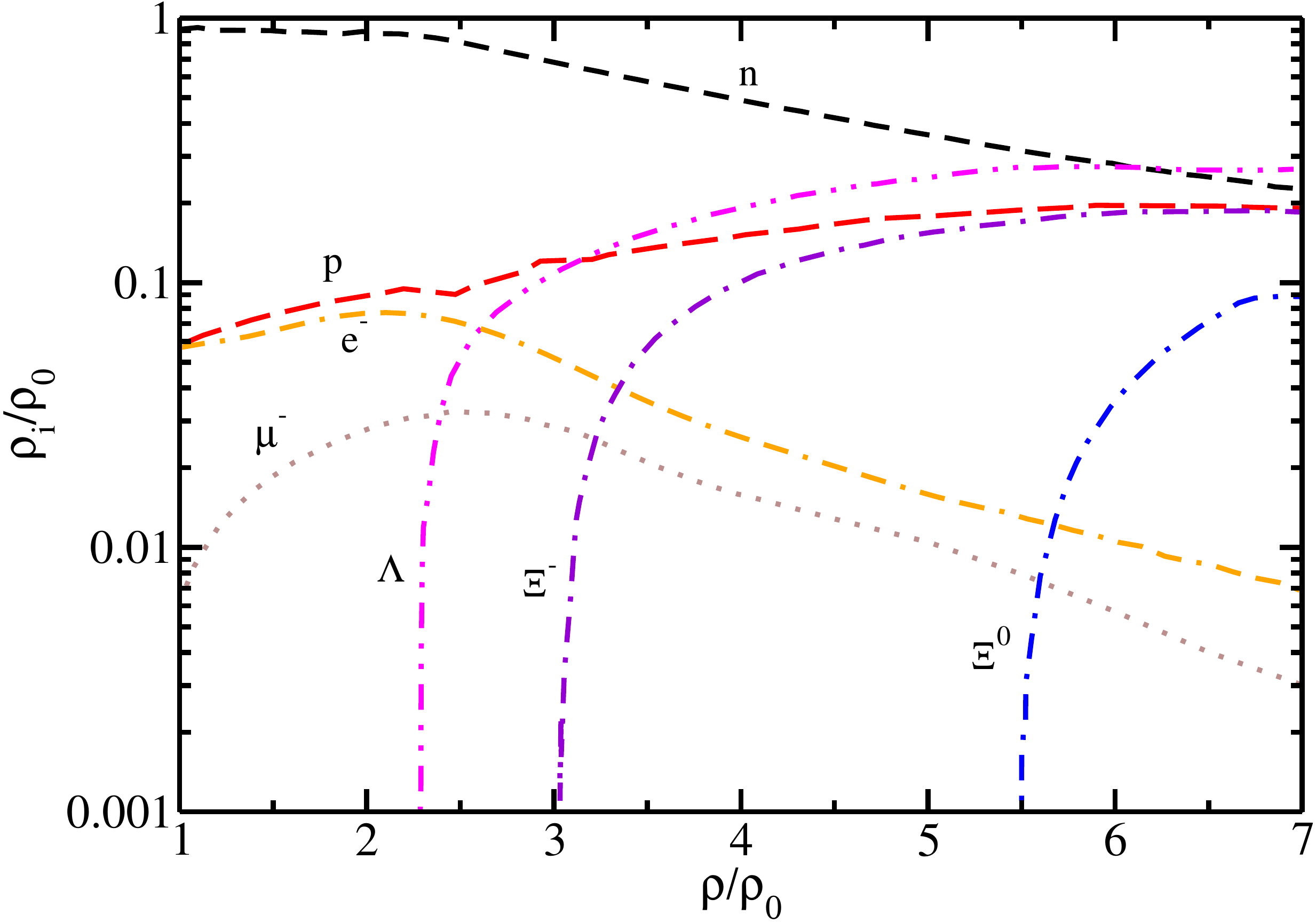}{0.5\textwidth}{(c)}
		\fig{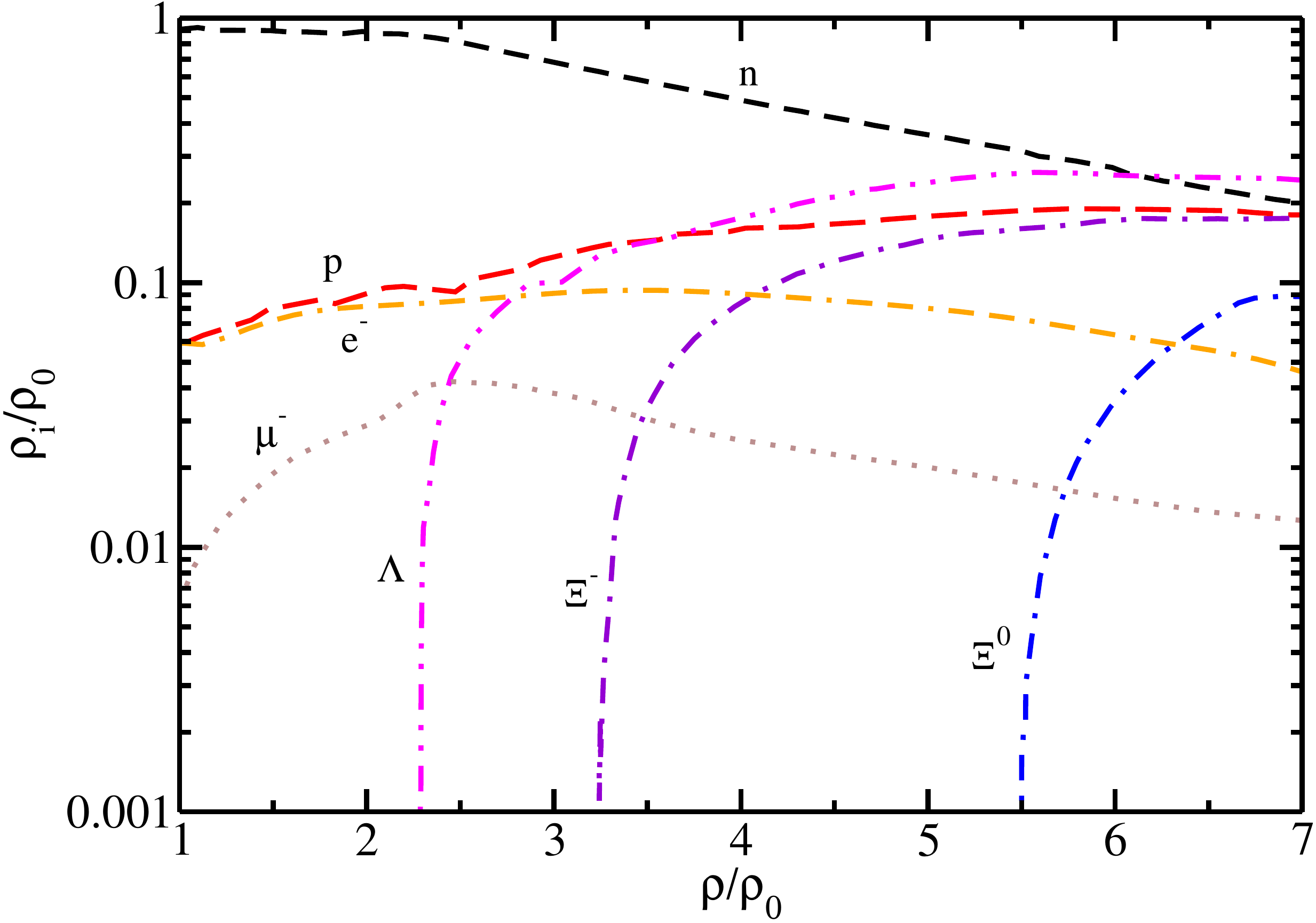}{0.5\textwidth}{(d)}
	}
\caption{Particle fraction of the baryons and leptons as a function of normalized baryon density for the DD-MEX model without magnetic field (a) and with magnetic field with different magnetic dipole moments b) $\mu$=5$\times$10$^{30}$ Am$^2$, c) $\mu$=5$\times$10$^{31}$ Am$^2$, and d) $\mu$=1$\times$10$^{32}$ Am$^2$.}
\label{fig6}
\end{figure*}

Figure \ref{fig5} shows the variation of the transverse pressure vs the energy density for the DD-MEX parameter set with and without hyperons. The solid line represents the variation in the pressure without including magnetic field ($\mu$=0), while the other lines represent the variation obtained with the magnetic field at magnetic dipole moments $\mu$=5$\times$10$^{30}$ Am$^2$, 5$\times$10$^{31}$ Am$^2$ and 10$^{30}$ Am$^2$. The insets show the pressure at higher values of the energy density. It is clear that the change in the pressure at a given value of energy density is larger for a NS with hyperons as compared to the NS without hyperons, which implies that the EoS with hyperons becomes stiffer than without hyperons (when compared to the $B$= 0 case) in the presence of strong magnetic field. The reason for such behavior will be discussed in the following. The magnetic field produced at the magnetic dipole moment 5$\times$10$^{30}$ Am$^2$ is of the order of 10$^{16}$ G at the center, which is small enough to be indistinguishable from the zero magnetic case. For higher magnetic dipole moments, the magnetic field produced $\approx$ 4 $\times$ 10$^{17}$ G is strong enough to increase the matter pressure to higher values, thus producing a distinguishable effect. 

Figure \ref{fig6} shows the particle fractions as a function of baryon density for beta stable NS matter obtained using the DD-MEX EoS. Figure \ref{fig6} panel (a) displays the fractions in the absence of magnetic field, while panels (b), (c), and (d) depict the particle fractions in the presence of magnetic fields with fixed magnetic dipole moments $\mu$=5$\times$10$^{30}$ Am$^2$, $\mu$=5$\times$10$^{31}$ Am$^2$, and $\mu$=5$\times$10$^{32}$ Am$^2$, respectively. Clearly, in all the cases, the $\Lambda$ particle is the dominant hyperonic component, which starts appearing in the density range 2 - 3 $\rho_0$ \citep{PhysRevC.81.035803,VIDANA2013367,rather2018role,doi:10.1142/S0218301318500970}. The neutral $\Xi$ hyperon appears at a density $\approx$ 5.5 $\rho_0$ for $B$=0, which remains unaltered with the inclusion of the magnetic field. As expected, the charged particles are more strongly affected by the magnetic field and an increase in their population is seen with the increase in the magnetic field strength. For $B$=0 (and all other cases), the $e^-$ and $\mu^-$ population is large at low densities, which suppresses the appearance of $\Xi^-$ hyperons. With an increase in the magnetic dipole moment, the magnetic field  strength increases, which shifts the appearance of $\Xi^-$ hyperon from $\approx$ 2.5 to around $\approx$ 3.5$\rho_0$. At this threshold, the density of the negatively charged leptons $e^-$ and $\mu^-$ starts to drop, as the charge neutrality condition from (\ref{eq13}) allows the $\Xi^-$ hyperon to take over. Similarly, the appearance of the $\Lambda$ hyperon accelerates the disappearance of neutrons, as both are neutral particles. Overall, we see that the addition of magnetic field increases the population of leptons (re-leptonizes) and correspondingly decreases the population of hyperons (de-hyperonizes), which renders the EoS stiffer \citep{Tolos_2016}.

Because of the repulsive nature of the $\Sigma$ potential, the formation of $\Sigma^0$ and $\Sigma^-$ is totally suppressed for the densities considered in the present work. The absence of $\Sigma$ hyperons is supported by the fact that no bound $\Sigma^-$ hypernuclei has been found yet, despite several searches \citep{HARADA2006206,HARADA2015312} and this has been reproduced in several works \citep{PhysRevC.64.025801,particles3040043,Guichon:2018uew}. The inclusion of strong magnetic fields do not change this feature. The properties of nucleonic and hyperonic stars at different values of the magnetic dipole moment, which correspond to different magnetic field values at the surface and the center, are shown in table \ref{tab4}.

\begin{figure}[t]
	\includegraphics[scale=0.34]{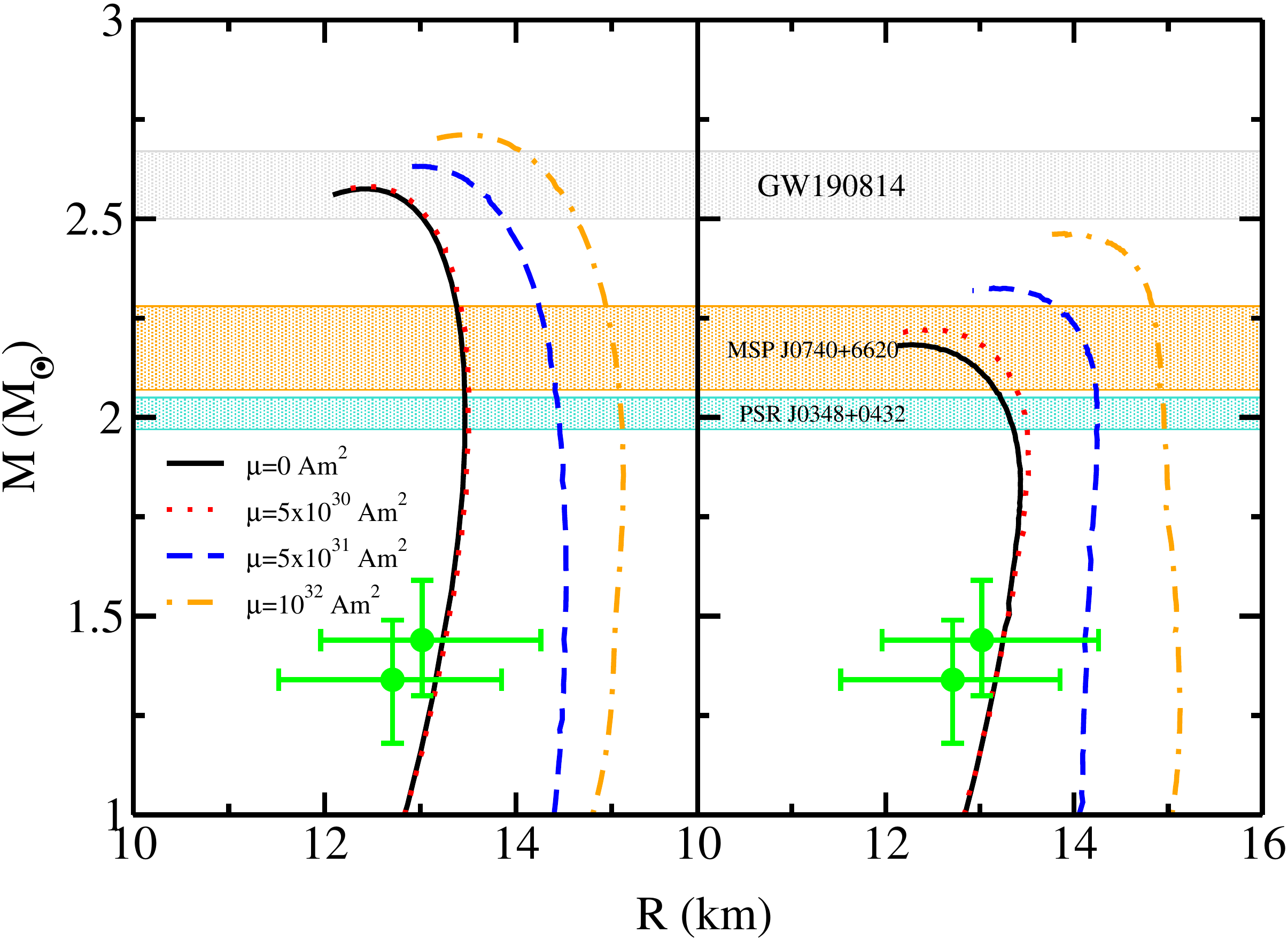}
	\caption{ Relation between mass and circumferential radius for a NS without magnetic field and with magnetic field effects considering different magnetic dipole moments without hyperons (left panel) and with hyperons (right panel) using the  DD-MEX parameter set. The colored areas show the recent constraints inferred from GW190814, MSP J0740+6620, and PSR J0348+0432 \citep{Abbott_2020a,Cromartie2020,Antoniadis1233232}. The constraints on the mass-radius limits inferred from  NICER  \citep{Miller_2019a,Riley_2019} are also shown.}
	\label{fig7} 
\end{figure}

The effect of magnetic field on the mass-radius relation of a NS with and without hyperons is displayed in figure \ref{fig7}. These calculations are performed for the DD-MEX EoS using the LORENE library \citep{LORENE}. Different values of magnetic  dipole moment are used to obtain different values of the magnetic field at the stellar surface and at the center. As can be seen, the NS maximum mass without hyperons increases from 2.575 $M_{\sun}$ for $B$=0 to 2.711 $M_{\sun}$ for $\mu$=10$^{32}$ Am$^2$. The corresponding radius changes from 12.465 to 13.474 km. The radius at 1.4$M_{\sun}$ increases by around 1 km. With the hyperons included, the mass increases from 2.183 to 2.463 $M_{\sun}$ when magnetic field effects are included with $\mu$=10$^{32}$ Am$^2$. The variation obtained in the mass-radius is larger for hyperonic stars than for the pure nucleonic stars due to the additional effect of de-hyperonization that takes place due to the magnetic field. The de-hyperonization results in the enhancement of the matter pressure $P_m$ for a given $\mathcal{E}_m$.

\begin{figure}[t]
	\includegraphics[scale=0.34]{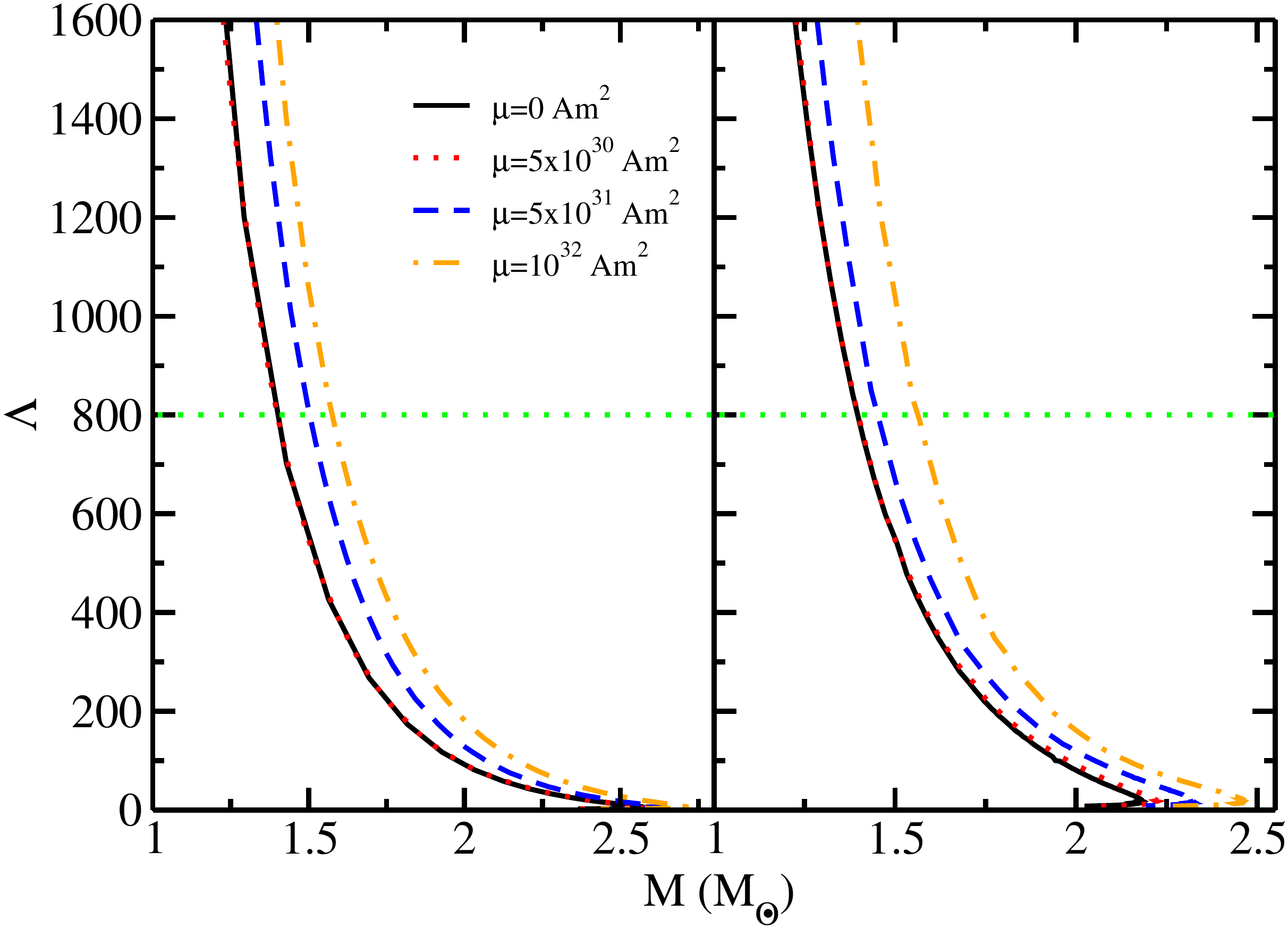}
	\caption{ Dimensionless tidal deformability ($\Lambda$) as a function of NS mass for EoS without magnetic field (solid line) and with magnetic field effects considering different magnetic dipole moments. Left panel shows the results obtained for NS without hyperons, while right panel shows the results for NS with hyperons.  The orange dotted line represents the upper limit on the dimensionless tidal deformabilty set by measurement from GW170817 \citep{PhysRevLett.119.161101}.}
	\label{fig8} 
\end{figure}

For higher magnetic fields produced at magnetic dipole moments $\mu$=5$\times$ 10$^{31}$ and $\mu$=10$^{32}$ Am$^2$, the effect of the magnetic field is seen to be very large at smaller stellar masses. For low magnetic fields, pure nucleonic stars still satisfy the possible maximum mass constraint from the GW190814 event, implying the possibility that of its secondary component to be a magnetar. The radius constraints inferred from NICER are satisfied by both nucleonic as well as hyperonic stars with low magnetic fields. For central magnetic fields $\approx$ 7 $\times$ 10$^{16}$ G, the EoS obtained for pure nucleonic matter satisfies the radius constraints from NICER measurement. For hyperonic matter, a lower magnetic field $\approx$ 4 $\times$ 10$^{16}$ G produces a hyperonic star with radius that satisfies NICER constraints. This is due to the fact that for magnetic fields less than 10$^{17}$ G, the deformation produced in the stellar structure is negligible and, hence, the variation in radius is too small when compared to the  non-magnetic case. Note that the surface magnetic field of the stars observed by NICER is expected to be $~\approx10^{8}$ G \citep{Miller_2019a}.\par 
Figure \ref{fig8} shows the variation in the dimensionless tidal deformability as a function of gravitational mass with magnetic field effects and considering different values of magnetic dipole moment. The results for EoS with and without hyperons are shown. For pure nucleonic stars, the tidal deformability increases to a value $\Lambda_{1.4} \approx$ 1500  for a central magnetic field  of 4.55$\times$ 10$^{17}$ G produced fixing the  magnetic dipole moment to $\mu$=10$^{32}$ Am$^2$, thus violating the constraint on the dimensionless tidal deformability from GW170817, which provides an upper limit of 800 on $\Lambda_{1.4}$ at 90\% confidence \citep{PhysRevLett.119.161101}.  Seen in table \ref{tab4}, a small magnetic field produces a NS with tidal deformability larger than the upper limit from GW170817. This confirms that the BNS merger event GW170817 did not consist of magnetars.

\begin{deluxetable}{p{1.5cm}|p{1.7cm}p{1.5cm}p{1.5cm}p{1.5cm}|
		p{1.7cm}p{1.5cm}p{1.5cm}p{1.5cm}}
	\tablenum{4}
	\tablecaption{\hspace{8.cm}{\bf{Table 4}}. \ Stellar properties: \ Maximum mass ($M_{max}$), maximum mass radius ($R_{max}$), canonical mass radius ($R_{1.4}$), and\newline 
	m\hspace{9.05cm} dimensionless tidal deformability ($\Lambda_{1.4}$) of pure nucleonic and hyperonic star for the DD-MEX EoS without magnetic field \newline 
	m\hspace{9.15cm}and with magnetic field effects for different values of the magnetic dipole moment.  \label{tab4}}
    \movetableright9.5cm	
	\tablewidth{0pt}
	\tablehead{
			\multirow{3}{1.5cm}{$\mu$ (Am$^2$)}
		&&\multicolumn{3}{p{2.5cm}|}{Nucleonic star} & %
		&\multicolumn{3}{p{2.5cm}}{Hyperonic star}\\
		\cline{2-9}
		&\parbox[t]{0.2cm}{\centering M$_{max}$ \\ ($M_{\sun}$) }  &\parbox[t]{0.2cm}{\centering R \\ (km) }&\parbox[t]{0.2cm}{\centering R$_{1.4}$  \\ (km) }&$\Lambda_{1.4}$&\parbox[t]{0.2cm}{\centering M$_{max}$ \\ ($M_{\sun}$) }  &\parbox[t]{0.2cm}{\centering R \\ (km) }&\parbox[t]{0.2cm}{\centering R$_{1.4}$  \\ (km) }&$\Lambda_{1.4}$ }
	\startdata 
	\hline
0 &2.575&12.465&13.168&791.483&2.183&12.238&13.168&791.483\\
5$\times$ 10$^{30}$  &2.580&12.536&13.235&802.801&2.224&12.506&13.168&791.483\\
5$\times$ 10$^{31}$ &2.632&13.024&14.507&1175.35&2.325&13.269&14.112&998.882\\
10$^{32}$ &2.711&13.474&15.027&1517.09&2.463&13.894&15.105&1559.194\\
	\enddata
\end{deluxetable}
For hyperonic stars, the value at 1.4$M_{\sun}$ remains unchanged when considering a magnetic dipole moment of $\mu$=5$\times$ 10$^{30}$ Am$^2$. As the magnetic dipole moment increases, stronger magnetic fields increase the stellar radius, which allows the tidal deformability $\Lambda_{1.4}$ to reach a value around 1550. For a nucleonic star and a hyperonic star with dimensionless tidal deformability well within the limit of GW170817 at 90\% confidence, a magnetic field with a maximum value of $\approx$ 2 $\times$ 10$^{16}$ G is required. See Ref~\citep{Biswas:2019gkw} for a discussion of the tidal deformability of deformed stars and Ref.~\citep{Zhu:2020imp} for a discussion on how very large magnetic fields generate new high-order corrections to the deformability that can modify the gravitational-wave evolution of magnetars. Note that magnetic NSs can also be a source of gravitational waves due to their asymmetric nature \citep{Sur:2020imd,Haskell:2007bh,Gomes:2019paw,Bonazzola:1995rb,Frieben:2012dz}.

\subsection{Additional hyperon couplings}
\label{sub3}
To investigate how different hyperon couplings and hyperon potentials affect the results we presented so far, we use a more general symmetry group SU(3) to determine the coupling constants of all baryons and mesons \citep{PhysRevC.89.025805,LOPES2021122171}. For the $\omega$ meson, we have 
\begin{equation}
\frac{g_{\omega \Lambda}}{g_{\omega N}} = \frac{4+2\alpha_v}{5+4\alpha_v},
\frac{g_{\omega \Sigma}}{g_{\omega N}} = \frac{8-2\alpha_v}{5+4\alpha_v},
\frac{g_{\omega \Xi}}{g_{\omega N}} = \frac{5-2\alpha_v}{5+4\alpha_v}.
\end{equation}
For the $\rho$ meson, we have
\begin{equation}
\frac{g_{\rho \Sigma}}{g_{\rho N}}=2\alpha_v,
\frac{g_{\rho \Xi}}{g_{\rho N}}=2\alpha_v-1,
\frac{g_{\rho \Lambda}}{g_{\rho N}}=0.
\end{equation}
The hyperon-scalar meson coupling constants are fixed so as to reproduce the following optical potentials \citep{PhysRevLett.67.2414,DOVER1984171,SCHAFFNER199435,PhysRevC.62.034311}
\begin{equation}
\begin{gathered}
U_{\Lambda}=-28 ~\mathrm{MeV},\\
U_{\Sigma}=+30 ~\mathrm{MeV},\\
U_{\Xi}=-18 ~\mathrm{MeV}.
\end{gathered}
\end{equation}
The coupling constants at different values of the parameter $\alpha_v$ are displayed in table \ref{newcoup}. For $\alpha_v$ = 1.0, the hybrid SU(6) group is recovered.

\begin{deluxetable}{p{1.4cm}|p{1.35cm}p{1.35cm}p{1.35cm}p{1.35cm}}[h!]
	\tablenum{5}
	\tablecaption{Hyperon meson coupling constants for different values of the hyperon coupling parameter $\alpha_v$. \label{newcoup}}
	\tablewidth{0pt}
	\tablehead{
		&$\alpha_v$ = 1.0&$\alpha_v$ = 0.75&$\alpha_v$ = 0.50&$\alpha_v$ = 0.25 }
	\startdata
	$g_{\omega \Lambda}/g_{\omega N}$&0.667&0.687&0.714&0.75\\
	$g_{\omega \Sigma}/g_{\omega N}$&0.667&0.812&1.00&1.25\\
	$g_{\omega \Xi}/g_{\omega N}$&0.333&0.437&0.571&0.75\\
	$g_{\rho \Sigma}/g_{\rho N}$&2.0&1.5&1.0&0.5\\
	$g_{\rho \Xi}/g_{\rho N}$&1.0&0.5&0.0&-0.5\\
	$g_{\sigma \Lambda}/g_{\sigma N}$&0.610&0.626&0.653&0.729\\
	$g_{\sigma \Sigma}/g_{\sigma N}$&0.403&0.514&0.658&0.850\\		
	$g_{\sigma \Xi}/g_{\sigma N}$&0.318&0.398&0.500&0.638\\
	\enddata
\end{deluxetable}

\begin{figure}[h]
	\includegraphics[scale=0.34]{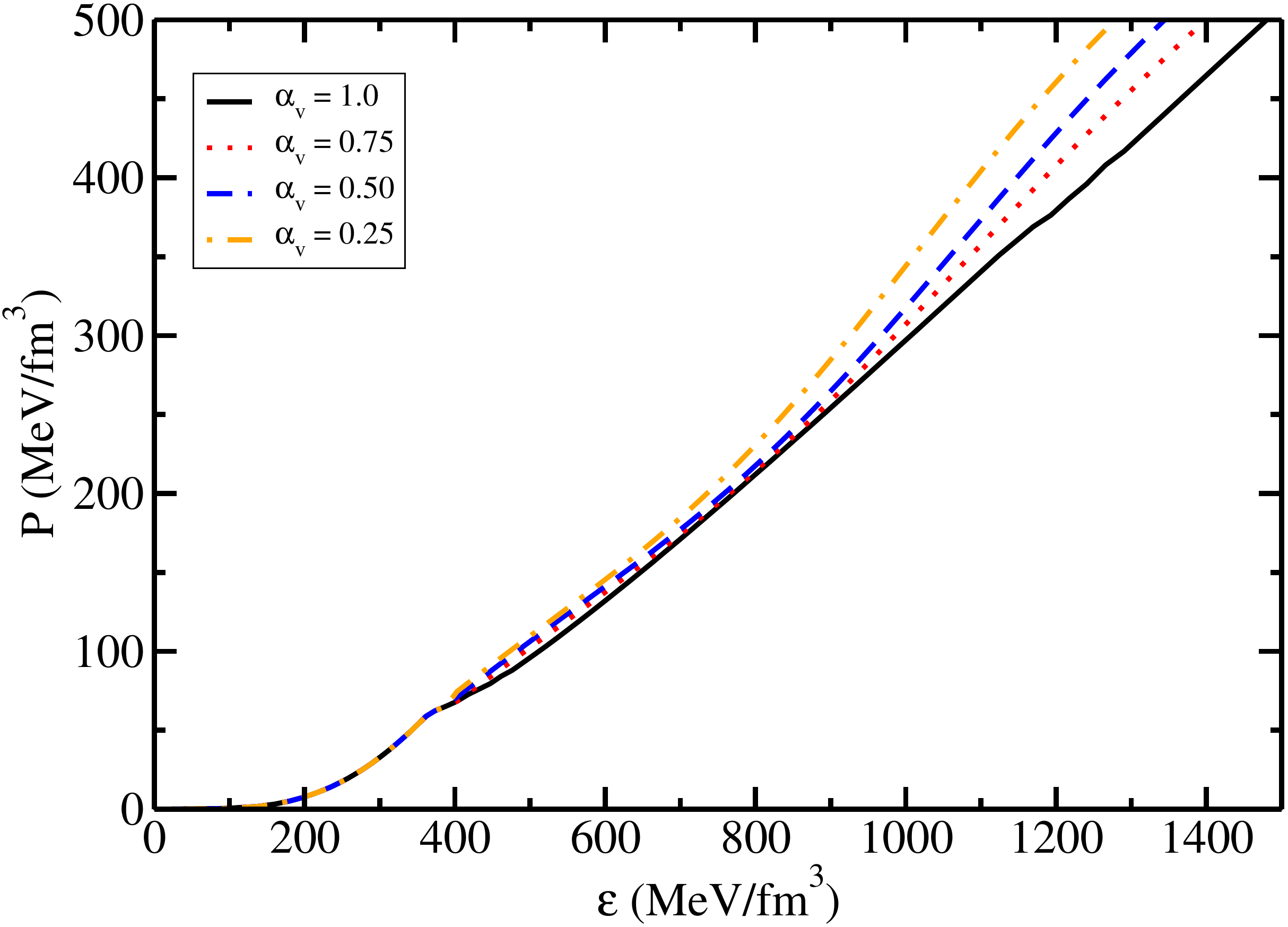}
	\caption{ EoS for the DD-MEX parameter set at different values of the parameter $\alpha_v$.  }
	\label{fig9} 
\end{figure}
The particle population of hyperons and other particles depend upon the hyperon meson coupling constants, which vary with the parameter $\alpha_v$.  Changing the value of $\alpha_v$ from 1.0 to 0.25 suppresses strange particles. The suppression of $\Xi^-$  increases the lepton fraction at lower value of $\alpha_v$. The neutrons and protons are the most populated particles at $\alpha_v$=0.25. 
Figure \ref{fig9} displays the EoS obtained using different values of the parameter $\alpha_v$. As the $\alpha_v$ value decreases from 1.0 to 0.25, the stiffness of the EoS increases due to the increase in the value of $\omega$ meson couplings.

\begin{figure}[t]
	\includegraphics[scale=0.34]{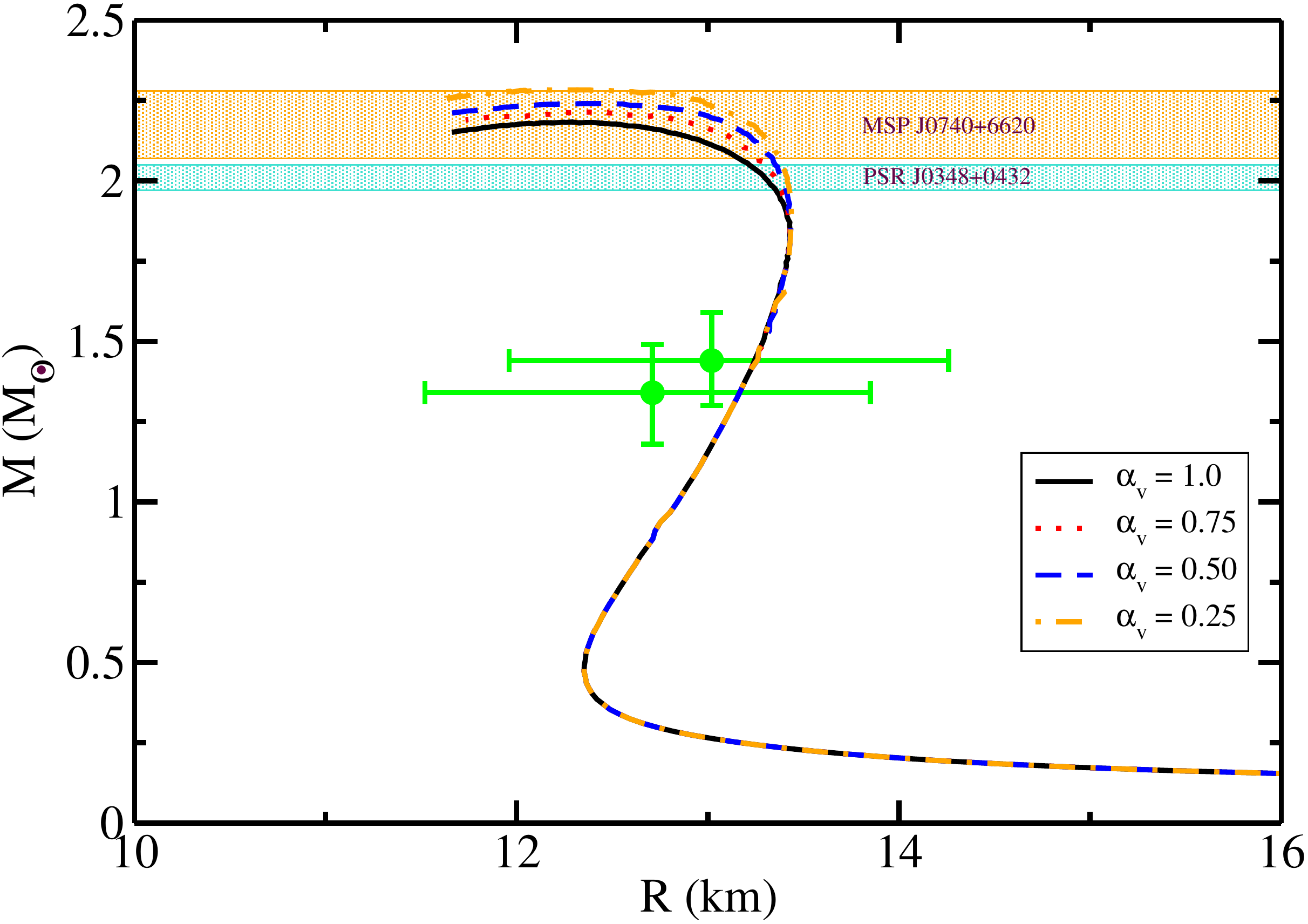}
	\caption{ Mass-Radius for a NS with hyperons at different values of the parameter $\alpha_v$ using the  DD-MEX parameter set. The shaded regions show the recent constraints inferred from GW190814, MSP J0740+6620, and PSR J0348+0432 \citep{Cromartie2020,Antoniadis1233232}. The green overlaid arrows show constraints on the mass-radius limits inferred from  NICER  \citep{Miller_2019a,Riley_2019}.}
	\label{fig10} 
\end{figure}

Figure \ref{fig10} displays the mass-radius relation for hyperonic stars without magnetic field effects at different values of $\alpha_v$. Lowering the value of $\alpha_v$  stiffens the EoS, which increases the maximum mass to 2.283 $M_{\sun}$ for $\alpha_v$=0.25 (from 2.183  for $\alpha_v$=1.0).  It is to be mentioned that the stellar properties of hyperonic stars obtained at $\alpha_v$ =1.0  almost resemble to that obtained from the hyperon potentials given by  (\ref{pot1}). The change in the hyperon potentials alters the values of sigma meson couplings by a small fraction and, hence, the changes obtained in the particle population and stellar properties are negligible.

\begin{figure}[h!]
	\includegraphics[scale=0.34]{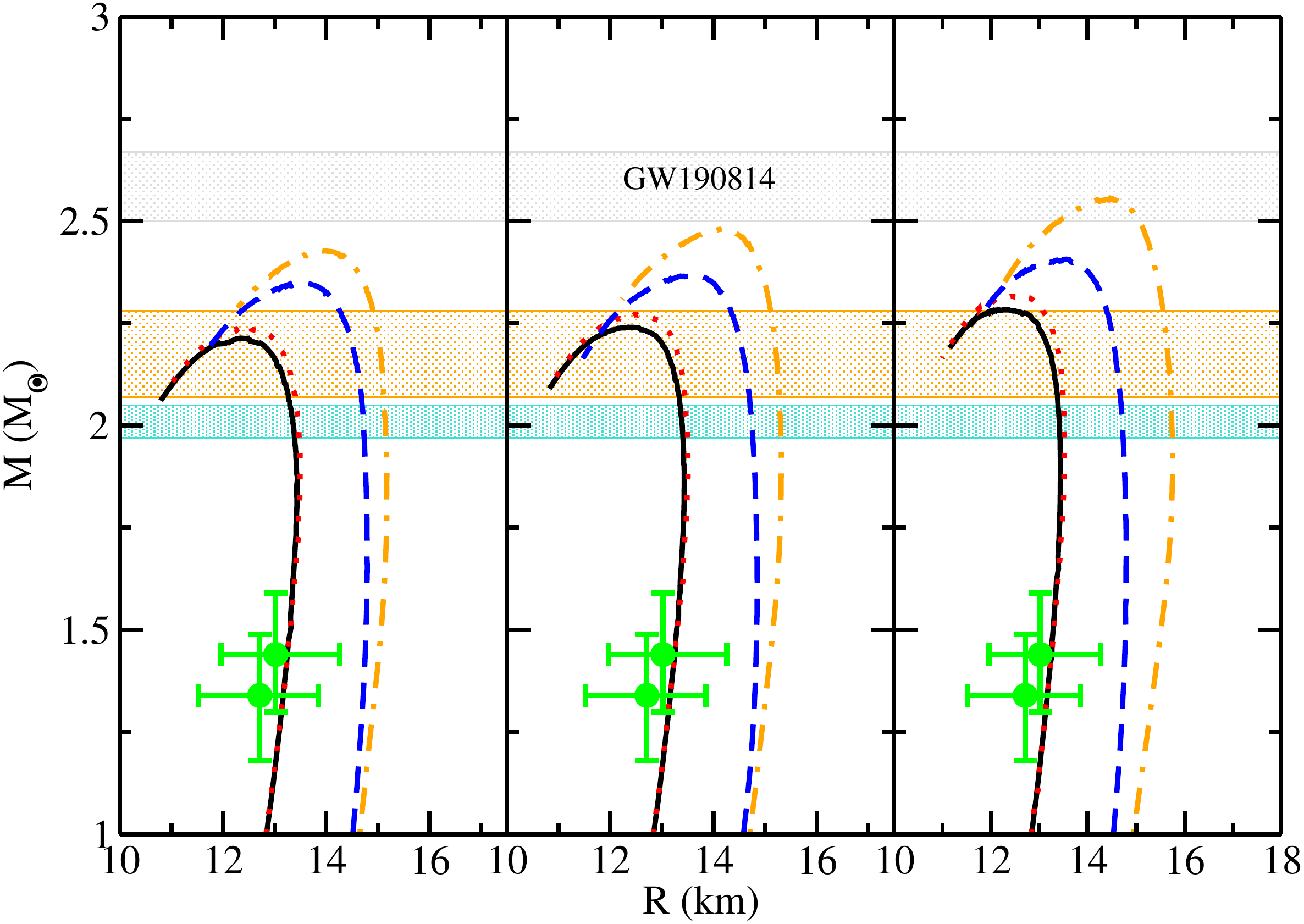}
	\caption{ Relation between mass and circumferential radius for a hyperonic star with magnetic field effects considering different magnetic dipole moments at different hyperon couplings  $\alpha_v$=0.75 (left panel),  $\alpha_v$=0.50 (middle panel) and  $\alpha_v$=0.25 (right panel) using the  DD-MEX parameter set. The colored areas show the recent constraints inferred from GW190814, MSP J0740+6620, and PSR J0348+0432 \citep{Abbott_2020a,Cromartie2020,Antoniadis1233232}. The constraints on the mass-radius limits inferred from  NICER  \citep{Miller_2019a,Riley_2019} are also shown.}
	\label{fig11} 
\end{figure}
With the addition of magnetic field effects, the results of different hyperon couplings on hyperonic EoSs is determined. Figure \ref{fig11} displays the relation between mass and radius for a hyperonic star at different $\alpha_v$ values without magnetic field and with magnetic field effects considered at different values of magnetic dipole moment. For $\alpha_v$=0.75, the maximum mass increases to a value 2.437 $M_{\sun}$ for a magnetic dipole moment $\mu$=10$^{32}$ Am$^2$, which corresponds to a central magnetic field of 3.77 $\times$ 10$^{17}$ G. Similarly, for $\alpha_v$=0.50 and 0.25, the maximum mass reaches a value 2.480 and 2.556$M_{\sun}$, respectively, at $\mu$=10$^{32}$ Am$^2$. This implies that the secondary component of GW190814 could be a hyperonic magnetar. For the present, hyperon couplings with a magnetic dipole moment $\mu$=5 $\times$ 10$^{31}$ Am$^2$, which corresponds to a central magnetic field greater than 10$^{17}$ G, the deviation in the hyperonic star radius at canonical mass is very large, around 1.5 km, as compared to that obtained from previous couplings. But with even stronger magnetic fields, the deviation obtained is less in the present case. This implies that a change in hyperon couplings effects the stellar properties, especially the radius at canonical mass at strong central magnetic field.

\begin{figure}[h!]
	\includegraphics[scale=0.34]{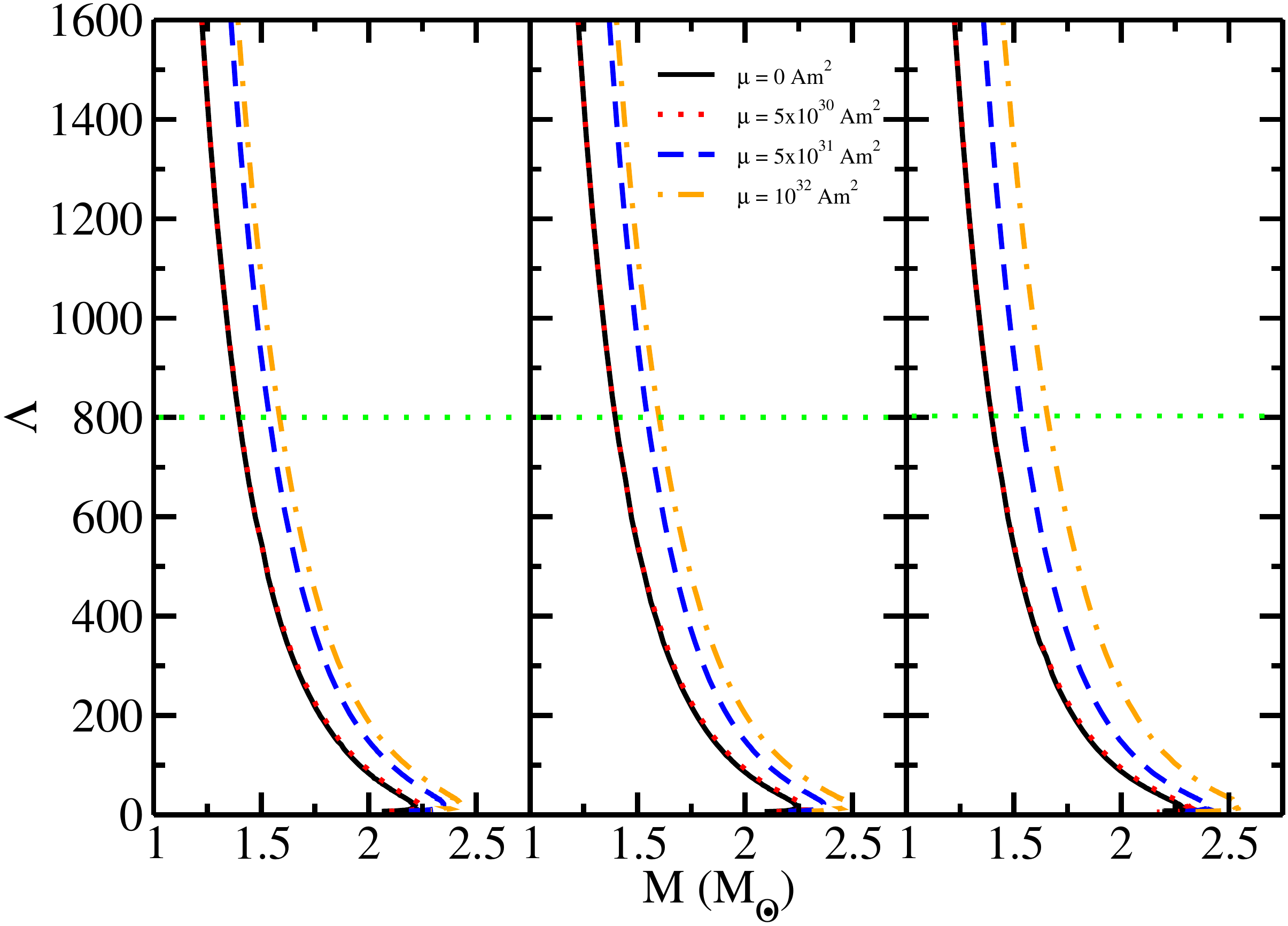}
	\caption{ Dimensionless tidal deformability  as a function of NS mass for EoSs without magnetic field (solid line) and with magnetic field effects considering different magnetic dipole moments. Different panels show the results obtained for hyperonic EoSs at different values of the parameter $\alpha_v$. The orange dotted line represents the upper limit on the dimensionless tidal deformability set by measurement from GW170817 \citep{PhysRevLett.119.161101}.}
	\label{fig12} 
\end{figure}

\begin{deluxetable}{p{1.1cm}|p{1.0cm}p{1.0cm}p{1.0cm}p{1.2cm}|
		p{1.0cm}p{1.0cm}p{1.0cm}p{1.2cm}|	p{1.0cm}p{1.0cm}p{1.0cm}p{1.2cm} }
	\tablenum{6}
	\tablecaption{\hspace{8.3cm}{\bf{Table 6}}. Stellar properties: Maximum mass ($M_{max}$), corresponding radius ($R_{max}$),  canonical mass radius ($R_{1.4}$), and dimensionless
	m\hspace{9.4cm}tidal deformability ($\Lambda_{1.4}$) of hyperonic star for the DD-MEX EoS at different values of the parameter $\alpha_v$ without magnetic field and
	m\hspace{9.4cm}with magnetic field with different values of the magnetic dipole moment.
		\label{tab6}}
		\movetableright9.7cm
	\tablewidth{0pt}
	\tablehead{
		\multirow{3}{0.2cm}{$\mu$ (Am$^2$)}&&\multicolumn{3}{p{2.5cm}|}{$\alpha_v$=0.75} & %
		&\multicolumn{3}{p{2.5cm}}{$\alpha_v$=0.50}
		&\multicolumn{3}{p{2.5cm}}{$\alpha_v$=0.25}\\
		\cline{2-13}
		&\parbox[t]{0.2cm}{\centering M$_{max}$ \\ ($M_{\sun}$) }  &\parbox[t]{0.2cm}{\centering R \\ (km) }&\parbox[t]{0.2cm}{\centering R$_{1.4}$  \\ (km) }&$\Lambda_{1.4}$&\parbox[t]{0.2cm}{\centering M$_{max}$ \\ ($M_{\sun}$) }  &\parbox[t]{0.2cm}{\centering R \\ (km) }&\parbox[t]{0.2cm}{\centering R$_{1.4}$  \\ (km) }&$\Lambda_{1.4}$&\parbox[t]{0.2cm}{\centering M$_{max}$ \\ ($M_{\sun}$) }  &\parbox[t]{0.2cm}{\centering R \\ (km) }&\parbox[t]{0.2cm}{\centering R$_{1.4}$  \\ (km) }&$\Lambda_{1.4}$}
	\startdata 
	\hline
		0  &2.217&12.335&13.168&791.483&2.239&12.335&13.168&791.483&2.283&12.453&13.168&791.483\\
	5$\times$ 10$^{30}$ 
	&2.240&12.543&13.213&815.398&2.271&12.632&13.224&869.281&2.316&12.652&13.237&914.012\\
	5$\times$ 10$^{31}$  &2.349&13.522&14.741&1338.714&2.366&13.708&14.821&1397.389&2.407&13.563&14.938&1425.185\\
	10$^{32}$  &2.437&13.995&14.992&1510.783&2.480&14.224&15.069&1615.288&2.556&14.469&15.403&1879.880\\
	\enddata
\end{deluxetable}

Figure \ref{fig12} displays the variation in the dimensionless tidal deformability of a hyperonic star with magnetic field effects considered at different values of magnetic dipole moment. For small values of the magnetic dipole moment at different values of $\alpha_v$, the $\Lambda$-M curve follows similar pattern as for the previous hyperon couplings. Since the variation in radius at canonical mass are large for the present couplings, the tidal deformability, $\Lambda_{1.4}$ obtained is also large. For $\alpha_v$=0.25, the dimensionless tidal deformability reaches a value of $\approx$ 1900 for magnetic dipole moment 10$^{32}$ Am$^2$. The properties of hyperonic stars using different hyperon coupling values according to the value of $\alpha_v$ with and without magnetic field effects considered at different magnetic dipole moments, are shown in table \ref{tab6}.

\section{Summary and Conclusion}
\label{summary}

We modeled massive nucleonic and hyperonic stars that fulfill the constraints set by the observation of the possibly most massive neutron star (NS) ever detected (in the secondary object of the gravitational wave event GW190814) using a density-dependent relativistic mean field model (DD-RMF). This simple, yet powerful formalism provides the freedom necessary to fulfill simultaneously nuclear and astrophysical constraints. The results obtained from the TOV equations show that the hyperons soften the EoS, lowering the maximum mass of NS to around 2.2$M_{\sun}$, thereby satisfying more conservative massive constraints from the astrophysical observations. The radius of the NS canonical mass is seen to be insensitive to the appearance of hyperons. Both the nucleonic and the hyperonic stars satisfy the constraints from mass-radius limits inferred from NICER observations and tidal deformability constraints from the LIGO and VIRGO collaborations.

We also studied the effects of strong magnetic fields on DD-RMF nucleonic and hyperonic EoSs. We investigated the EoS and particle populations using a realistic chemical potential-dependent magnetic field that was developed by solving Einstein-Maxwell equations. For very strong magnetic fields, the spherical symmetric solutions obtained by solving the TOV equations lead to an overestimation of the mass and the radius and, hence, cannot be used for determining stellar properties. For this reason, we used the LORENE library to determine stellar properties of magnetic NSs. For low values of the magnetic dipole moment, implying lower strengths of magnetic fields, the EoS resembles the non-magnetic one. For higher magnetic dipole moments, the EoS stiffens at higher energy density. The amount of stiffness is larger in case of hyperonic EoSs than for the case of pure nucleonic ones. As the magnetic field strength is increased, the particle fractions of leptons ($e^-$ and $\mu^-$) increases at higher densities and the appearance of charged hyperons ($\Xi^-$) is delayed.

The stiffening in the EoS caused by the changes in population described above  increases the maximum mass of magnetic stars. For a small dipole moment of 5$\times$ 10$^{30}$ Am$^2$, the nucleonic and hyperonic mass-radius profiles resemble the non-magnetic case due to the lower magnetic field produced. For higher magnetic dipole moments, although the variation in the maximum mass is small, a variation of about 1 km is seen in the radius of the NS canonical mass. For hyperonic stars, the maximum mass increases by $\approx$ 0.3 $M_{\sun}$, allowing them to satisfy the GW190814 possible mass constraint for a dipole moment $\mu\ge10^{32}$ Am$^2$ when different coupling schemes are considered. For strong magnetic fields, the different coupling schemes for hyperons predict increases in radius of $\approx0.2-1$ km. Thus, we see that different hyperon couplings and different hyperon potentials  populate stellar matter differently and, hence, can change stellar properties significantly.

The change in the dimensionless tidal deformability for NS was studied. It was found that, for higher value of the magnetic dipole moment, the tidal deformability of the canonical mass surpasses the upper limit of 800 set by the measurement from GW1701817 at 90\% confidence, which is consistent with the acknowledgement of such objects possessing weak magnetic fields. No such measurement of tidal deformability is available for GW190814, as it has no tidal signatures. But if in the future, measurements of tidal deformability of massive NSs such as the one measured for the secondary object in GW190814 are $\le$ 800, this would imply that object could not be a hyperonic magnetar.\par 
In the future, we will investigate the effect of considering more exotic particles in our modelling of magnetic NSs. These include, kaons, $\Delta$-resonances and deconfined quark matter (with and without mixtures of phases). We also intend to analyze combined effects of magnetic fields and fast rotation on stars containing exotic matter, as both of these aspects could be important in young stars \citep{Lasky:2019fdv,Yu:2019tyy,Lasky:2017hff}.

\begin{acknowledgments}
	AAU acknowledges the Inter-University Centre for Astronomy and Astrophysics, Pune, India for support via an associateship and for hospitality. VD thanks Valentina Sarti and Nicolas Yunes for useful discussions and acknowledges support from the National Science Foundation under grant PHY-1748621 and PHAROS (COST Action CA 16214).
\end{acknowledgments}

\appendix
\section{}
\label{A}
\numberwithin{equation}{section}
Within the relativistic density-dependent mean field (DD-RMF) model, the equation of motion for baryons obtained by applying the Euler Lagrange equations to the Lagrangian density (\ref{lag}) is written as
\begin{equation}
\sum_{B}\Bigg[i\gamma_{\mu}D^{\mu}-\gamma^0 \Bigg(g_{\omega}(\rho_B)\omega +\frac{1}{2}g_{\rho}(\rho_B)\rho \tau_3 +\sum_R (\rho_B)\Bigg)-M_{B}^*\Bigg]\psi_{B}=0,
\end{equation} 
where, $M_B^* =M_B-g_{\sigma}(\rho_B)\sigma$  is the effective mass of baryons and $\sum_R$ is the rearrangment term due to the density dependence of the coupling constants
\begin{equation}
\sum_R(\rho_B) = -\frac{\partial g_{\sigma}}{\partial \rho_B}\sigma \rho_s +\frac{\partial g_{\omega}}{\partial \rho_B}\omega \rho_B+\frac{1}{2}\frac{\partial g_{\rho}}{\partial \rho_B}\rho \rho_3.
\end{equation}
The equation of motion for meson fields are
\begin{align}
m_{\sigma}^2 \sigma &= g_{\sigma}(\rho_B)\rho_s, \nonumber \\
m_{\omega}^2 \omega &= g_{\omega}(\rho_B)\rho_B,\nonumber  \\
m_{\rho}^2 \rho &= \frac{g_{\rho}(\rho_B)}{2}\rho_3,
\end{align} 

where the scalar density $\rho_s$, baryon density $\rho_B$, and isovector densities $\rho_{s3}$, and $\rho_3$, are given as
\begin{align}
\rho_s &= \sum_{B}\bar{\psi}\psi, \nonumber \\
\rho_B &= \sum_{B}\psi^{\dagger}\psi, \nonumber\\ 
\rho_{s3}& = \sum_{B}\bar{\psi}\tau_3\psi, \nonumber\\
\rho_3 &= \sum_{\alpha}\psi^{\dagger}\tau_3\psi =\rho_p -\rho_n.
\end{align}

With the above equations, the expression for the energy density and pressure are
\begin{align} \label{eq18}
\mathcal{E}_m &= \sum_{B} \frac{2}{(2\pi)^3}\int_{0}^{k_B} d^3 k E_B^* (k) + \frac{1}{2}m_{\sigma}^2 \sigma^2-\frac{1}{2}m_{\omega}^2 \omega^2-\frac{1}{2}m_{\rho}^2 \rho^2 \nonumber \\
&+g_{\omega}(\rho_B)\omega \rho_B+\frac{g_{\rho}(\rho_B)}{2}\rho \rho_3,  \nonumber \\
P_m&= \sum_{B} \frac{2}{3(2\pi)^3}\int_{0}^{k_B} d^3 k \frac{k^2}{E_B^* (k)} -\frac{1}{2}m_{\sigma}^2 \sigma^2+\frac{1}{2}m_{\omega}^2 \omega^2 \nonumber \\
&+\frac{1}{2}m_{\rho}^2 \rho^2-\rho_B \sum_R (\rho_B),
\end{align}
where, $E_{B}^*=\sqrt{k_{B}^2+M_B^{*2}}$. The rearrangment term $\sum_{R}(\rho_B)$ contributes to the pressure only.

In the presence of magnetic field, the scalar and vector density for charged baryons $cb$, uncharged baryons $ub$, and leptons follow as \citep{Broderick_2000}
\begin{align}
\rho_s^{cb}&=\frac{|q_{cb}|B M_{cb}^{*2}}{2\pi^2}\sum_{\nu=0}^{\nu_{max}}r_{\nu}  ln\Bigg(\frac{k_{F,\nu}^{cb}+E_F^{cb}}{\sqrt{M_{cb}^{*2}+2\nu|q_{cb}|B}}\Bigg), \nonumber \\
\rho_s^{ub}&=\frac{M_{ub}^{*2}}{2\pi^2}\Bigg[E_F^{ub} k_F^{ub} -  M_{ub}^{*2} ln\Bigg(\frac{k_{F,\nu}^{ub}+E_F^{ub}}{M_{ub}}\Bigg)\Bigg],  \nonumber \\
\rho^{cb}&= \frac{|q_{cb}|B}{2\pi^2}\sum_{\nu=0}^{\nu_{max}}r_{\nu} k_{F,\nu}^{cb}, \nonumber \\
\rho^{ub}&=\frac{(k_F^{ub})^3}{3\pi^2},\nonumber \\
\rho_l &=\frac{|q_l|B}{2\pi^2}\sum_{\nu=0}^{\nu_{max}}r_{\nu} k_{F,\nu}^l,
\end{align}
where $r_{\nu}$ is the Landau degeneracy of $\nu$ level.
%\begin{equation}
%r_{\nu}= \begin{cases}
%1 & \text{for $\nu$=0 }\\
%2 & \text{for $\nu \ne$ 0}
%\end{cases}  
%\end{equation}
The expressions for the baryon and lepton energy densities in the presence of magnetic field become
\begin{align}
\mathcal{E}_{cb}&= \frac{|q_{cb}|B}{4\pi^2}\sum_{\nu=0}^{\nu_{max}}r_{\nu} \nonumber \\
&\times \Bigg[k_{F,\nu}^{cb}E_F^{cb}+(M_{cb}^{*2}+2\nu|q_{cb}|B) ln\Bigg(\frac{k_{F,\nu}^{cb}+E_F^{cb}}{\sqrt{M_{cb}^{*2}+2\nu|q_{cb}|B}}\Bigg)\Bigg], \nonumber \\
\mathcal{E}_{ub} &= \frac{1}{8\pi^2}\Bigg[k_{F}^{ub}(E_F^{ub})^3+(k_F^{ub})^3E_F^{ub} -M_{ub}^{*4} ln \Bigg(\frac{k_{F}^{ub}+E_F^{ub}}{M_{ub}^*}\Bigg)\Bigg],\nonumber \\
\mathcal{E}_{l}&= \frac{|q_l|B}{4\pi^2}\sum_{\nu=0}^{\nu_{max}}r_{\nu} \nonumber \\
&\times \Bigg[k_{F,\nu}^{l}E_F^{l}+(m_{l}^{2}+2\nu|q_l|B) ln\Bigg(\frac{k_{F,\nu}^{l}+E_F^{l}}{\sqrt{m_{l}^2+2\nu|q_l|B}}\Bigg)\Bigg].
\end{align}


%\bibliography{ref}{}
%\bibliographystyle{aasjournal}


%\begin{appendix}

\end{document}